\documentclass[
    aps,                % American Physical Society
    prx,                % PRX style
    reprint,            % Two-column, final look
    superscriptaddress, % Author affiliations as superscripts
    nofootinbib         % Put footnotes at bottom of page, not in bib
]{revtex4-2}

\usepackage{graphicx}

\usepackage{xcolor}
\usepackage{xr-hyper}
\usepackage{hyperref}
\usepackage{xr}
\usepackage{amsmath}

\begin{document}

\title{Mechanisms of Spatiotemporal Damage Evolution in Double Polymer Networks}
\author{M. Le Goff}
\affiliation{Univ.~Grenoble Alpes, CNRS, LIPhy, 38000 Grenoble, France}
\affiliation{Institut fur Theoretische Physik, Universitat Innsbruck, 6020 Innsbruck, Austria}
\author{L. Ortellado}
\affiliation{Univ.~Grenoble Alpes, CNRS, LIPhy, 38000 Grenoble, France}
%\altaffiliation{Current address: Some other place, Othert\"own,Germany}
\author{J. Tian}
\affiliation{Institute of Nuclear Physics and Chemistry, China Academy of Engineering Physics, 621999 Mianyang, China}
\author{M. Bouzid}
\affiliation{Univ.~Grenoble Alpes, CNRS, 3SR, 38000 Grenoble, France}
\author{J-L. Barrat}
\author{K. Martens}
\affiliation
{Univ.~Grenoble Alpes, CNRS, LIPhy, 38000 Grenoble, France}

\begin{abstract}
Double polymer networks exhibit a striking enhancement of toughness compared to single networks, yet the microscopic mechanisms governing stress redistribution, damage evolution, and fracture remain incompletely understood. Using large-scale coarse-grained molecular dynamics simulations under uniaxial deformation, we resolve bond scission statistics, local stress redistribution following individual bond-breaking events, and the spatiotemporal evolution of damage in single- and double-network architectures. We show that while the early mechanical response is dominated by the pre-stretched sacrificial network, damage evolution in double networks follows a qualitatively distinct pathway. In contrast to single networks, where anisotropic stress redistribution promotes rapid localization and catastrophic fracture, the presence of a soft matrix in double networks induces a screening of stress redistribution generated by sacrificial bond scission. This screening suppresses correlated rupture events and stabilizes multiple damage zones, leading to a strongly delocalized damage landscape over a broad deformation range. At larger strains, when the matrix becomes load-bearing, damage progressively localizes, ultimately triggering fracture. By isolating the dynamics of individual damage zones, we further demonstrate that matrix-mediated stress screening stabilizes defects and delays localization. Together, these results identify stress-screening–induced damage delocalization as a central microscopic mechanism underlying toughness enhancement in multiple-network elastomers.
\end{abstract}

\maketitle

\section{Introduction}

% General context and problematic
Multi-component soft solids are composed of two or more intertwined polymer networks, whose assembly and inter-network synergy enable control over visco-elasto-plastic properties ~\cite{li2024design,creton201750th,gong2010double,mugnai2025interspecies,burla2019stress}.
Under external loading, when stresses exceed a critical threshold, the microstructure irreversibly reorganizes across multiple scales to relax mechanical constraints~\cite{zhao2014multi}. Depending on how this reorganization proceeds, materials may fracture abruptly or accommodate extensive plastic deformation prior to failure. Mechanical response is commonly characterized by the elastic stiffness at small deformations and the fracture energy (toughness) at large deformations, whose simultaneous enhancement remains a central challenge in soft solids.

%%%%%%%%%%%%%%%%%%%%%%%%%%%%%%%%%%%%%%%%%%%

Double-network (DN) materials have emerged as a prototypical class of soft solids that combine high stiffness and enhanced toughness compared to single networks~\cite{gong2003double,nakajima2009true,Ducrot2014,Webber2007,Slootman2022}. They consist of two interpenetrating polymer networks with contrasting mechanical roles: a highly cross-linked, brittle sacrificial network embedded within a weakly cross-linked, highly stretchable matrix network.

DNs can sustain large deformations and extensive damage without propagating macroscopic fracture, even when pre-notched, and the high fracture energies reported experimentally have been attributed to energy dissipation by bond scission near the crack tip~\cite{Ducrot2014,gong2010double}. Despite extensive efforts to engineer tough hydrogels and elastomers by tuning network architecture and preparation protocols~\cite{gong2003double}, a detailed molecular understanding of the mechanisms governing the large increase in fracture energy remains incomplete~\cite{gong2010double,creton201750th}.

%Increased toughness and load sharing
The increase in toughness in multiple interpenetrating networks appears to be a ubiquitous phenomenon, observed across a wide range of materials, from hydrogels~\cite{gong2003double}, elastomers~\cite{Ducrot2014}, to macroscopic networks~\cite{king2019macroscale}, suggesting a generic mechanistic origin of the enhancement of mechanical properties~\cite{nakajima2017generalization,zhao2014multi,creton201750th}.
%namely the ``sacrificial bond principle" \cite{nakajima2017generalization}.

The phenomenological models developed by Brown \cite{brown2007model} and Tanaka \cite{tanaka2007local} rationalize DN toughness within a common process-zone framework: above a critical stress, the sacrificial network undergoes irreversible damage (multiple cracking/fragmentation), while the matrix bridges the damaged region and carries load. A finite damaged zone forms around the crack tip, and macroscopic fracture proceeds within this zone. As a result, the measured fracture energy is dominated by the work required to create and extend the damaged region, rather than by the intrinsic surface energy of either network. While this picture successfully explains large toughness and weak rate dependence, it remains largely phenomenological: damage is introduced through threshold parameters and an effective softened modulus, without an explicit constitutive coupling between scission kinetics, load transfer between networks, and the strongly nonlinear (finite-extensibility) stress–strain responses that govern strain localization and the spatial distribution of dissipation.

Mechanophore experiments on notched double and triple network elastomers support the interpretation that fracture proceeds through extensive sacrificial bond scission and the formation of a damaged process zone ahead of the crack tip \cite{Ducrot2014}. In unnotched DN, this behavior gives rise to a necking regime in which the matrix predominantly carries the load after extensive rupture of the sacrificial network, leading to macroscopically inhomogeneous deformation \cite{na2006necking,Millereau2018}. The onset of this regime depends on the molar concentrations of both networks as well as their topological architecture \cite{fukao2020effect,lu2024tensile}.

Recent coarse-grained molecular simulations by Higuchi \emph{et al.}~\cite{higuchi2018fracture} confirmed the existence of a two-stage fracture process in double networks obtained without swelling, by carefully tuning the topology of both networks. This picture was further supported by simulations of swollen double networks by Tauber \emph{et al.}~\cite{tauber2021sharing}, who demonstrated that both the global stress response and the microscopic fracture mechanics are governed by internetwork interactions, leading to a redistribution of stress that strongly deviates from affine predictions. 

Complementary athermal two-dimensional spring-network simulations by Walker and Fielding~\cite{walker2025toughness} showed that strong coupling between the two networks can reduce the effective transmission of stress between breaking sacrificial bonds, thereby suppressing crack-like avalanches and promoting diffuse damage. While this work highlights an important limiting mechanism by which internetwork coupling can inhibit localization, the delocalization observed in that model is imposed through an effective interaction kernel and does not resolve the dynamical route by which damage becomes spatially distributed in realistic three-dimensional polymer networks.

While damage evolution in single and double networks has been investigated experimentally on larger scales \cite{Millereau2018, orr2025probing}, conclusive spatially resolved analyses via simulations have remained limited \cite{higuchi2018fracture}, and the mechanisms governing damage delocalization and localization are still not fully understood.

Here we identify a microscopic mechanism governing damage evolution and toughness in double polymer networks, which has not been explicitly resolved at the level of stress redistribution and damage dynamics. We show that the presence of a soft matrix induces a screening of the local stress redistribution generated by sacrificial bond scission, even at small strains where the matrix is not yet macroscopically load-bearing. This screening suppresses the anisotropic stress amplification characteristic of single networks, inhibits correlated bond-breaking events, and stabilizes multiple damage zones. As a result, damage accumulates in a spatially delocalized manner long before load sharing becomes dominant, delaying the emergence of a single fracture path and enabling enhanced energy dissipation.

Using large-scale particle-based simulations, the present work therefore pursues two main objectives:
(i) to systematically characterize how damage evolution, spatial delocalization, and toughness depend on the pre-stretch $\lambda_0$, and
(ii) to identify the mesoscale mechanisms, such as dissipation pathways, stress screening, and defect stabilization, that collectively underpin the enhanced toughness and delayed localization observed in double networks.

By performing large-strain uniaxial deformation of single- and double-network systems, we reveal that damage propagation in double networks follows a qualitatively distinct spatiotemporal pathway compared to single networks. While the early mechanical response is controlled by the sacrificial network, as anticipated from previous simulation studies~\cite{tauber2021sharing}, damage in single networks rapidly concentrates into a single region, leading to macroscopic fracture at moderate strains. In contrast, double networks initially develop multiple spatially separated nanovoids within the pre-stretched sacrificial network, while the soft matrix remains intact. Damage accumulates within these regions without immediate coalescence, resulting in a strongly delocalized damage landscape over a broad deformation range.

To elucidate the microscopic origin of this behavior, we analyze energy dissipation and stress redistribution at the level of individual bond-breaking events. We show that, compared to single networks, sacrificial bond rupture in double networks generates partially screened stress perturbations due to the presence of the matrix. Even at early stages, when the matrix is not yet macroscopically load-bearing, this screening suppresses the strongly anisotropic stress redistribution characteristic of single networks, thereby inhibiting correlated bond breaking and delaying the concentration of damage into a single fracture zone.

We further demonstrate this mechanism through a controlled study of an isolated damage island. In double networks, matrix-mediated stress relaxation around the defect markedly slows its growth and suppresses the rapid localization observed in single networks, providing direct evidence that inter-network coupling stabilizes distributed damage and hinders early damage concentration into a single crack.

\section{Polymer network model and stretching protocol} \label{sec:Model}

%%%%%%%%%%%%%%%%%%%%%%%%%%%%%%%%%%%%%%%%%%%%%%%%%%%%%%%%%%%%%%%%%%%%%%%%%%%%%%%%%%%%%%%%%%%%%%%%%%%%%%%%%%%%
%%%%%%%%%%%%%%%%%%%%%%%%%%%%%%  FIGURE 1 %%%%%%%%%%%%%%%%%%%%%%%%%%%%%%%%%%%%%%%%%%%%%%%%%%%%%%%%%%%%%%%%%%%
%%%%%%%%%%%%%%%%%%%%%%%%%%%%%%%%%%%%%%%%%%%%%%%%%%%%%%%%%%%%%%%%%%%%%%%%%%%%%%%%%%%%%%%%%%%%%%%%%%%%%%%%%%%%
\begin{figure*}[htp!]
  \includegraphics[width=\textwidth]{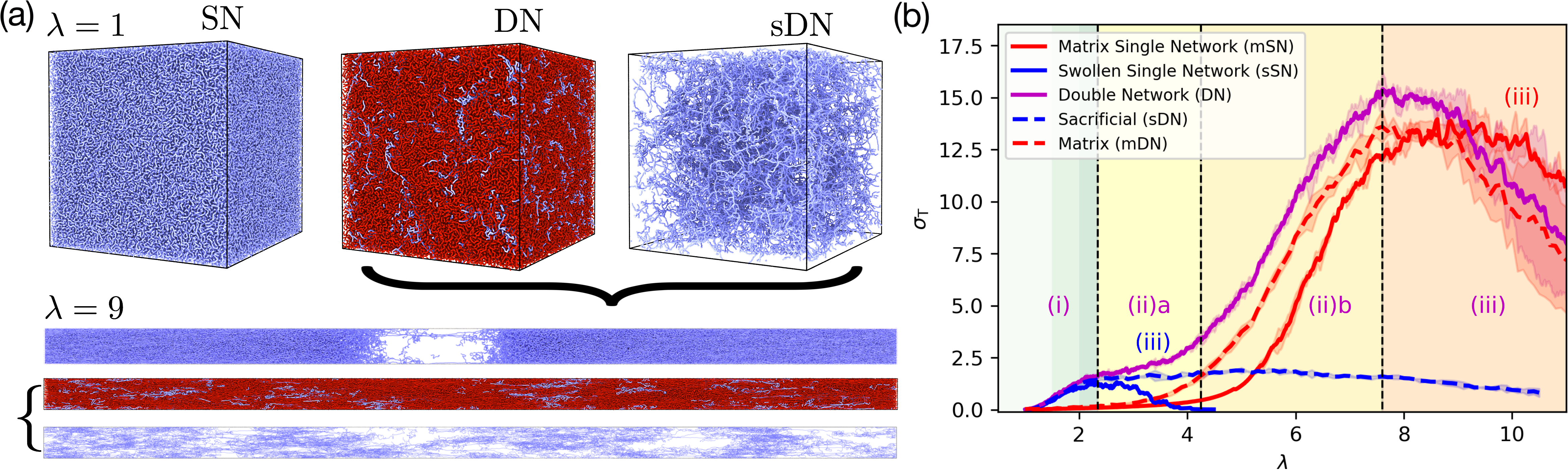}
  \caption{(a) Simulation snapshots for SN and DN samples (red and blue beads for the matrix and sacrificial network respectively).
  Top panel: initial configurations ($\lambda = 1.0$) for the SN (left panel), and the DN (showing both networks (middle panel) and only sacrificial network (right panel))
  Bottom panel: stretched configuration ($\lambda = 9.0$) for the SN (top panel), and the DN showing both networks (top) and sacrificial network only (bottom).
  (b) True stress as a function of stretch for a single (swollen) sacrificial network (sSN) with $\lambda_0=2.00$ (blue solid line), a matrix single network (mSN) (red solid line), a double network (DN) with $\lambda_0=2.00$ (magenta solid line) and the contributions of the sacrificial (sDN) (blue dashed line) and matrix (mDN) (red dashed line) networks within the DN. Different deformation regimes are shown using different colors: regime (i) extends up to the stress peak of the sSN (green); regime (ii)a starts after the stress peak (yellow) and is followed by regime (ii)b when the stress contribution of the mDN dominates over the sDN; regime (iii) starts after the stress peak (orange shaded area for DN, indicated by blue text for sSN). }
  \label{fgr:0}
\end{figure*}
%%%%%%%%%%%%%%%%%%%%%%%%%%%%%%%%%%%%%%%%%%%%%%%%%%%%%%%%%%%%%%%%%%%%%%%%%%%%%%%%%%%%%%%%%%%%%%%%%%%%%%%%%%%%
%%%%%%%%%%%%%%%%%%%%%%%%%%%%%%%%%%%%%%%%%%%%%%%%%%%%%%%%%%%%%%%%%%%%%%%%%%%%%%%%%%%%%%%%%%%%%%%%%%%%%%%%%%%%
%%%%%%%%%%%%%%%%%%%%%%%%%%%%%%%%%%%%%%%%%%%%%%%%%%%%%%%%%%%%%%%%%%%%%%%%%%%%%%%%%%%%%%%%%%%%%%%%%%%%%%%%%%%%

To study the response of polymer networks under uniaxial stretching, we employ a coarse-grained bead-spring model of the Kremer-Grest type~\cite{orr2025probing}, in which polymer strands are represented as chains of interacting connected beads. Excluded-volume interactions between nonbonded beads are described by a purely repulsive short-range potential, defining the bead radius. In the following all lengths are expressed in units of this radius. Beads are connected into chains via nonlinear elastic bonds, while damage is modeled through irreversible bond scission when the bond length exceeds a given distance. Finite chain flexibility is accounted for through an angular potential that controls bending rigidity. This minimal yet physically grounded model captures the essential ingredients required to investigate stress redistribution, damage accumulation, and fracture in polymer networks, while remaining computationally efficient for large-scale simulations. Full details of the interaction potentials and parameters are provided in Appendix~\ref{app:model-details}.

Disordered single- and double-network polymer systems (Fig.~\ref{fgr:1}(a)) are generated using a computational synthesis protocol inspired by experimental elastomer preparation and radical polymerization schemes \cite{perez2008polymer,tian2025influence} (Fig.~\ref{fgr:A1}(a)). 
Single networks are formed by polymerizing and crosslinking monomers into a randomly connected network at fixed density. 
Double networks are constructed sequentially by first swelling the initial network to a prescribed swelling ratio, thereby introducing an isotropic pre-stretch of its strands, and subsequently polymerizing a second, more weakly cross-linked network within the swollen matrix. This procedure results in two interpenetrated networks with distinct connectivities and mechanical roles: a pre-stretched sacrificial network and a highly extensible matrix network. The resulting structures reproduce key statistical features of experimental double networks, including strand-length distributions (Fig.~SI1) and controlled pre-stretch of the sacrificial network (Fig.~\ref{fgr:A1}(b)). Full details of the synthesis protocol and parameters are provided in Appendix~\ref{app:model-details}.

Mechanical response is probed by subjecting the systems to large-strain uniaxial extension under controlled conditions. Deformation is applied at a low constant rate (rate independent mechanical response regime) along a single axis (Fig.~\ref{fgr:0}(a)), with affine rescaling of beads positions followed by relaxation, allowing the material to respond through elastic deformation, plastic rearrangements, and bond scission. Lateral dimensions are maintained at constant pressure, enabling volume relaxation and preventing artificial transverse stresses. Periodic boundary conditions are used in all directions. The macroscopic stress response is computed from the virial stress tensor, and the true tensile stress is defined by subtracting the transverse contributions. This protocol provides access to the full nonlinear stress–strain behavior (Fig.~SI2) and damage evolution up to fracture. Simulation details and numerical parameters are provided in Appendix~\ref{app:model-details}.

%%%%%%%%%%%%%%%%%%%%%%%%%%%%%%%%%%%%%%%%%%%%%%%%%%%%%%%%%%%%%%%%%%%%%%%%
%%%%%%%%%%%%%%%%%%%%%%%%%%%%%%% RESULTS PART 1 %%%%%%%%%%%%%%%%%%%%%%%%%
\section{Three distinct deformation regimes in double networks} \label{sec:Results-1}
%%%%%%%%%%%%%%%%%%%%%%%%%%%%%%%%%%%%%%%%%%%%%%%%%%%%%%%%%%%%%%%%%%%%%%%%
%%%%%%%%%%%%%%%%%%%%%%%%%%%%%%%%%%%%%%%%%%%%%%%%%%%%%%%%%%%%%%%%%%%%%%%%

\begin{figure}[htp!]
  \includegraphics[width=\columnwidth]{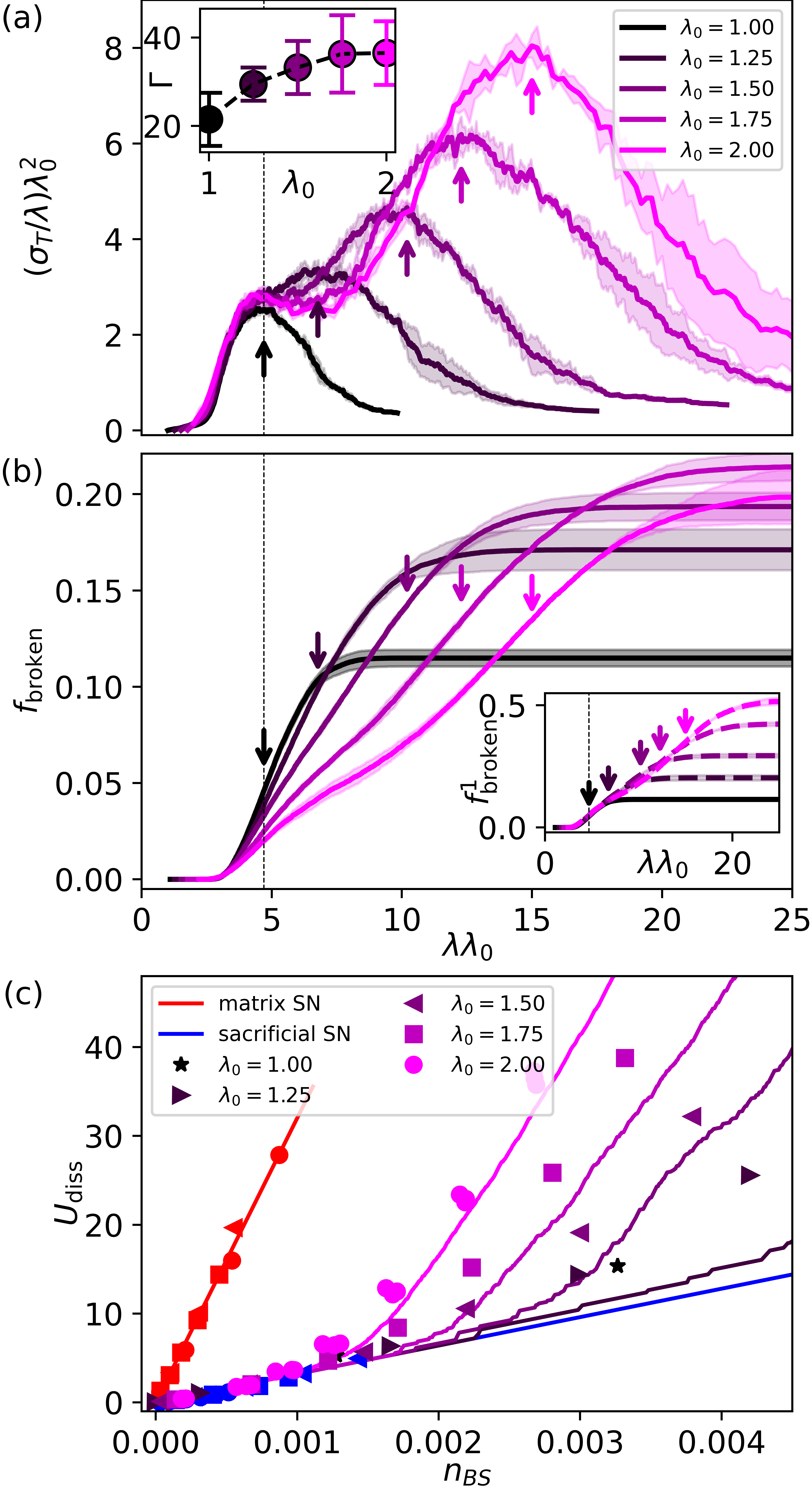}
  \caption{(a) Rescaled engineering stress as a function of the rescaled stretch $\lambda\lambda_0$ for different values of the isotropic pre-stretch $\lambda_0$. All data are averaged over three independent network realizations, and the light shaded areas indicate the corresponding error bars. We mark the early loading regime (i), which collapses for all curves in rescaled units, as well as the distinct fracture regimes (iii). The vertical arrows correspond to the position of the stress maximum, which marks the end of regime (ii) for each value of $\lambda_0$. The inset shows the fracture toughness $\Gamma$ (see Supplementary Information, Fig~SI5) as a function of the pre-stretch $\lambda_0$. 
  (b) Fraction of broken strands $f_\mathrm{broken}$ as a function of $\lambda\lambda_0$. Inset : Fraction of broken sacrificial strands $f_\mathrm{broken}^1$ as a function of $\lambda\lambda_0$. 
  (c) Dissipated energy density $U_\mathrm{diss}$ as a function of broken strands density $n_{BS}$ in regimes (i) and (ii) (see Supplementary Information, Fig~SI5). Red and blue solid lines represent linear fits to matrix and sacrificial single networks, respectively. Different symbols represent different pre-stretch values $\lambda_0$ for the DN (and the corresponding SNs are obtained by deleting the second network after synthesis of the DN).}
  \label{fgr:1}
\end{figure}

The mechanical response of double networks separates into three distinct regimes (depicted in the stress-stretch curve of Fig.~\ref{fgr:0}(b) for $\lambda_0=2.00$).
%%% Regime (i) %%%%%%%%%%%%%%%%%%%
Regime (i) in Fig.~\ref{fgr:0}(b) corresponds to the early loading stage, where the stress–stretch curves of the sacrificial (swollen) single network (sSN) and the double network (DN) collapse, indicating that the response is dominated by the stiff, pre-stretched sacrificial network (Fig.~\ref{fgr:1}(a)). 
The strands of the sacrificial network exhibit a strain-hardening response already at small stretch values, due to the initial pre-stretch ($\lambda_0=2.00$) (hence there is no visible linear elastic regime for the sSN and DN in Fig.~\ref{fgr:0}(a)).
This initial strain hardening regime is followed by strain softening associated with damage, where up to 5\% of the sacrificial strands are broken (Fig.~\ref{fgr:1}(b)). 
Here, bond breaking occurs only in the sacrificial network (see Fig.~\ref{fgr:1}(b) and Fig.~\ref{fgr:A2}), and strongly correlates with the strands contour length and initial stretch (Fig.~\ref{fgr:A3}), suggesting that damage dynamics is dominated by the structure of the sacrificial network in regime (i) \cite{zhang2025fracture}.

%%% Regime (ii)a %%%%%%%%%%%%%%%%%%%
At larger stretch in the plateau-like sub-regime (ii)a, the load is carried predominantly by the sacrificial network within the double network (sDN), while the matrix network (mDN) remains weakly loaded.
Damage accumulates in the sDN through the formation of spatially distributed damage islands, rather than localizing into a single critical region.
As a result, the sacrificial network in the DN sustains load well beyond the fracture point of the corresponding single network (sSN), highlighting the stabilizing role of inter-network coupling.

%%% Regime (ii)b %%%%%%%%%%%%%%%%%%%
Sub-regime (ii)b is marked by a renewed increase in stress as mechanically coupled sacrificial and matrix strands jointly bear the load (Fig.~\ref{fgr:0}(b)).
Here, steric inter-network interactions are maximal, leading to a significant increase in the load carried by the mDN while maintaining a sustained load on the sDN (Fig.~SI3 and Fig.~SI4(b)).
Damage begins to accumulate in the matrix network (Fig.~\ref{fgr:A2}), while the sacrificial network exhibits a second maximum in its bond-breaking rate (Fig.~SI4(a)), even though the corresponding sSN has already fractured.
This coexistence of sacrificial and matrix damage leads to a two-step fracture scenario (separated by a decrease in rupture rate, see Fig.~\ref{fgr:1}(b)), in which sacrificial bond scission is prolonged and enhanced through inter-network coupling, consistent with previous observations \cite{tauber2021sharing}.

Crucially, however, the defining feature of this regime is not load sharing per se, but the persistence of a highly delocalized damage landscape. Multiple damage zones remain active simultaneously throughout the sample (Fig.~\ref{fgr:0}(a), bottom panel), reflecting the fragmentation of both the sacrificial and matrix networks into damage islands. This spatial fragmentation effectively screens stress redistribution between damage sites, suppressing the amplification of local stress concentrations that would otherwise trigger rapid crack localization. As a result, the double network is prevented from collapsing into a single dominant fracture path at this stage. Instead, damage progresses through the parallel growth and interaction of multiple micro-damage regions, substantially delaying catastrophic failure. In this sense, the creation and stabilization of the damage zones, and the associated inhibition of early localization, plays a more central role in enhancing toughness than load transfer alone.

%%% Regime (iii) %%%%%%%%%%%%%%%%%%%
Finally, regime (iii) corresponds to the fracture regime, in which damage (occurring predominantly in the mDN) concentrates into a single dominant region. 
Stress rapidly drops after a maximum as the matrix fails in zones where the sacrificial network is already severely degraded, leading to macroscopic crack formation.

%%%%%%%%%%%%%%%%%%%%%%%%%%%%%%%%%%%%%%%%%%%%%%%%%%%%%%%%%%%%%%%%%%%%%%%%%%%%%%%%%%%%%%%%%%%%%%%%%%%%%%%%%%%%
%%%%%%%%%%%%%%%%%%%%%%%%%%%%%%  FIGURE 2 %%%%%%%%%%%%%%%%%%%%%%%%%%%%%%%%%%%%%%%%%%%%%%%%%%%%%%%%%%%%%%%%%%%
%%%%%%%%%%%%%%%%%%%%%%%%%%%%%%%%%%%%%%%%%%%%%%%%%%%%%%%%%%%%%%%%%%%%%%%%%%%%%%%%%%%%%%%%%%%%%%%%%%%%%%%%%%%%
%%%%%%%%%%%%%%%%%%%%%%%%%%%%%%%%%%%%%%%%%%%%%%%%%%%%%%%%%%%%%%%%%%%%%%%%%%%%%%%%%%%%%%%%%%%%%%%%%%%%%%%%%%%%
%%%%%%%%%%%%%%%%%%%%%%%%%%%%%%%%%%%%%%%%%%%%%%%%%%%%%%%%%%%%%%%%%%%%%%%%%%%%%%%%%%%%%%%%%%%%%%%%%%%%%%%%%%%%
%%%%%%%%%%%%%%%%%%%%%%%%%%%%%%%%%%%%%%%%%%%%%%%%%%%%%%%%%%%%%%%%%%%%%%%%%%%%%%%%%%%%%%%%%%%%%%%%%%%%%%%%%%%%

%%%%%%%%%%%%%%%%%%%%%%%%%%%%%%%%%%%%%%%%%%%%%%%%%%%%%%%%%%%%%%%%%%%%%%%%%%%%%%%%%%%%%%%%%%%%%%%%%%%%%%%%%%%%
%%%%%%%%%%%%%%%%%%%%%%%%%%%%%%%%%%%% RESULTS PART 2 %%%%%%%%%%%%%%%%%%%%%%%%%%%%%%%%%%%%%%%%%%%%%%%%%%%%%%%%
%%%%%%%%%%%%%%%%%%%%%%%%%%%%%%%%%%%%%%%%%%%%%%%%%%%%%%%%%%%%%%%%%%%%%%%%%%%%%%%%%%%%%%%%%%%%%%%%%%%%%%%%%%%%
\section{Increase in toughness with pre-stretch}\label{sec:Results-2}
%%%%%%%%%%%%%%%%%%%%%%%%%%%%%%%%%%%%%%%%%%%%%%%%%%%%%%%%%%%%%%%%%%%%%%%%%%%%%%%%%%%%%%%%%%%%%%%%%%%%%%%%%%%%
%%%%%%%%%%%%%%%%%%%%%%%%%%%%%%%%%%%%%%%%%%%%%%%%%%%%%%%%%%%%%%%%%%%%%%%%%%%%%%%%%%%%%%%%%%%%%%%%%%%%%%%%%%%%

\begin{figure*}[ht]
  \includegraphics[width=\textwidth]{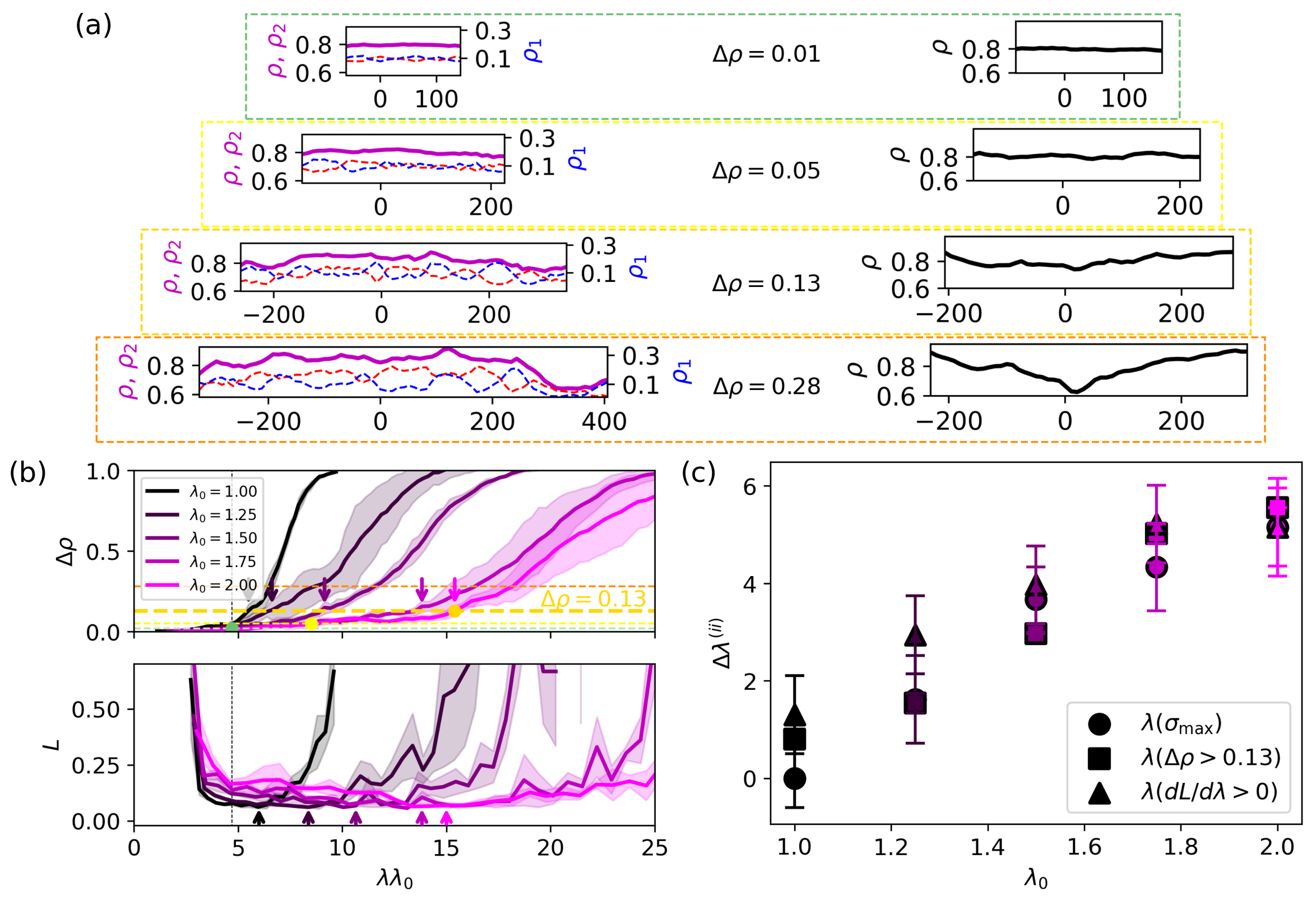}
  \caption{(a) Left panel: Profiles of density along the stretching ($x$) direction for the double network at the end of regime (i) (top panel), at the end of regime (ii)a (second panel) and the end of regime (ii)b (third panel) and in regime (iii) (bottom panel), as indicated by the colored horizontal dashed lines and dots in panel (b). Density profiles are shown for the full DN (magenta solid line), the sacrificial network (dashed blue line) and the matrix network (dashed red line). 
  Right panel: Density profile in the SN exhibiting the same level of localization $\Delta \rho = \rho_\mathrm{max}-\rho_\mathrm{min}$ as in the corresponding DN plot (as indicated by the dashed horizontal lines in panel (b)). 
  (b) Top panel: Difference $\Delta \rho$ between the maximum and minimum density along the stretching direction (averaged over the orthogonal direction $y$ and $z$ and using binned profiles with a bin size $\Delta x = 10$ monomeric size units) vs rescaled stretch $\lambda \lambda_0$. 
  Bottom panel: Localization parameter $L$ (see text) computed using a stretch window $\Delta \lambda = 0.4$ (40 strain steps) and a bin size $\Delta x = 20$ monomeric size units.
  (c) Extension of regime (ii) (in non rescaled stretch units) as a function of the swelling ratio $\lambda_0$. Regime (ii) starts at the end of regime (i) (Fig.~\ref{fgr:1}(a)) ($\lambda \lambda_0 \simeq 4.7$) and the stretch at which it ends is measured either from the position of the stress maximum as indicated by the arrows on Fig.~\ref{fgr:1}(a) (circular symbols), from the stretch at which $\Delta \rho$ becomes larger than 0.13 (squared symbols, see thick dashed line and arrows in panel (c)) or from the onset of bond breaking localization (where $dL/d\lambda>0$, see arrows on panel (b)) (triangular symbols).}
  \label{fgr:2}
\end{figure*}

%%% FIG 2A : Mechanical response for different pre-stretch values
The responses to uniaxial deformation of samples prepared with different values of the pre-stretch $\lambda_0$ are depicted in Fig.~\ref{fgr:1}(a).
% Regime (i)
The stress–stretch curves collapse when rescaled by the pre-stretch $\lambda_0$, confirming that
the macroscopic response in regime (i) is dominated by the stiff, pre-stretched sacrificial network.
% Regime (ii)
Regime (ii) becomes increasingly pronounced as $\lambda_0$ increases: the plateau-like sub-regime (ii)a extends over a broader stretch range (Fig.~\ref{fgr:2}(c)), reflecting the fact that highly pre-stretched sacrificial strands carry more load and accumulate damage more gradually (Fig.~SI4), allowing the matrix to delay damage localization over a wider strain interval. 
% Regime (iii)
As a consequence, both the maximum stress and the maximum stretch attained before fracture systematically increase with $\lambda_0$. 

%%% FIG 2A - inset : Increase in toughness with pre-stretch
The inset in Fig.~\ref{fgr:1}(a) shows that the fracture toughness $\Gamma$ rises with pre-stretch $\lambda_0$, capturing the enhanced ability of double networks to dissipate energy (see Supplementary Information, Fig.~SI5, for details on toughness measurements).
This increase in toughness correlates directly with the greater extent of sacrificial bond breaking observed in Fig.~\ref{fgr:1}(b), where networks with larger $\lambda_0$ exhibit a higher fraction of broken strands in the sacrificial network prior to failure. Thus, the growing prominence of regime (ii), the higher stress peak, and the increase in toughness all share a common microscopic origin in the enhanced damage accumulation enabled by sacrificial-network pre-stretching.

%%%%%%%%%%%%%%%%%%%%%%%%%%%%%%%%%%%%%%%%%%%%%%%%%%%%%%%%%%%%%%%%%%%%%%%%%%%%%%%%%%%%%%%%%%%%%%%%%%%%%%%%%%%%
%%%%%%%%%%%%%%%%%%%%%%%%%%%%%%  FIGURE 3 %%%%%%%%%%%%%%%%%%%%%%%%%%%%%%%%%%%%%%%%%%%%%%%%%%%%%%%%%%%%%%%%%%%
%%%%%%%%%%%%%%%%%%%%%%%%%%%%%%%%%%%%%%%%%%%%%%%%%%%%%%%%%%%%%%%%%%%%%%%%%%%%%%%%%%%%%%%%%%%%%%%%%%%%%%%%%%%%
%%%%%%%%%%%%%%%%%%%%%%%%%%%%%%%%%%%%%%%%%%%%%%%%%%%%%%%%%%%%%%%%%%%%%%%%%%%%%%%%%%%%%%%%%%%%%%%%%%%%%%%%%%%%
%%%%%%%%%%%%%%%%%%%%%%%%%%%%%%%%%%%%%%%%%%%%%%%%%%%%%%%%%%%%%%%%%%%%%%%%%%%%%%%%%%%%%%%%%%%%%%%%%%%%%%%%%%%%
%%%%%%%%%%%%%%%%%%%%%%%%%%%%%%%%%%%%%%%%%%%%%%%%%%%%%%%%%%%%%%%%%%%%%%%%%%%%%%%%%%%%%%%%%%%%%%%%%%%%%%%%%%%%

%%% FIG 2 c : Dissipated energy in double networks %%%%%
Fig.~\ref{fgr:1}(c) depicts the dissipated energy density $U_\mathrm{diss}$ (equivalent to the toughness $\Gamma$ when measured at the stress peak) as a function of the density of broken strands, $n_{BS}$, in regimes (i) and (ii) (we discarded the data in the fracture regime (iii), see Fig.~SI5).
For both sacrificial and matrix single networks (which differ only in terms of crosslink concentration $c_1 = 5 c_2$), $U_\mathrm{diss}^\mathrm{SN}$ grows linearly with $n_{BS}$ (for $n_{BS}$ sufficiently small), consistent with a Lake–Thomas–type mean-field argument, in which the dissipated energy scales with the total number of broken strands, with slopes $u_1$ and $u_2$, for the matrix and sacrificial networks respectively.
The large dissipated energy $u_2$ ($u_2 \simeq 10 u_1$) in the mSN may arise from non local energy dissipation mechanisms as shown to occur in disordered entangled polymer networks \cite{zheng2022fracture, zhang2025fracture}.

A simple Lake-Thomas argument suggests that the dissipated energy density could be written as $U_{\mathrm{diss}}^{\mathrm{DN}} = u_1 n_1 + u_2 n_2$, where $u_1$ and $u_2$ denote the energy densities dissipated per broken bond in the sacrificial and matrix networks, respectively.
Using the values of $u_1$ and $u_2$ obtained from the linear fits in Fig.~\ref{fgr:1}(c) (with $u_2 \simeq 10 u_1$), enables us to explain qualitatively the increase in toughness in double networks as $\lambda_0$ increases, as resulting from the additive contributions of broken bonds in each network. Unsurprisingly, this simple form underestimates the dissipated energy since the fit of the single network data are obtained at low stretch (and bond breaking dissipates more energy as the strand stretch increases, see section ~\ref{sec:Results-4}). Crucially, however, this energetic argument alone does not explain the mechanism by which double networks accommodate a substantially larger number of bond-breaking events without undergoing early damage localization and catastrophic fracture.

%%%%%%%%%%%%%%%%%%%%%%%%%%%%%%%%%%%%%%%%%%%%%%%%%%%%%%%%%%%%%%%%%%%%%%%%%%%%%%%%%%%%%%%%%%%%%%%%%%%%%%%%%%%%
%%%%%%%%%%%%%%%%%%%%%%%%%%%%%%%%%%%% RESULTS PART 3 %%%%%%%%%%%%%%%%%%%%%%%%%%%%%%%%%%%%%%%%%%%%%%%%%%%%%%%%
%%%%%%%%%%%%%%%%%%%%%%%%%%%%%%%%%%%%%%%%%%%%%%%%%%%%%%%%%%%%%%%%%%%%%%%%%%%%%%%%%%%%%%%%%%%%%%%%%%%%%%%%%%%%
\section{Relation between pre-stretch and delocalized damage}\label{sec:Results-3}
%%%%%%%%%%%%%%%%%%%%%%%%%%%%%%%%%%%%%%%%%%%%%%%%%%%%%%%%%%%%%%%%%%%%%%%%%%%%%%%%%%%%%%%%%%%%%%%%%%%%%%%%%%%%%%%%%%%%%%%%%%%%%%%%%%%%%%%%%%%%%%%%%%%%%%%%%%%%%%%%%%%%%%%%%%%%%%%%%%%%%%%%%%%%%%%%%%%%%%%%%%%%%%%%%%%%%%%%

%%%%%%%% FIGURE 3a : DENSITY PROFILES %%%%%%%%%%%%%%%%%%%%
Although the overall increase in toughness can be simply attributed to the larger number of broken bonds in double networks (DNs) as the pre-stretch $\lambda_0$ is increased, the key open question is how this increase in damage actually develops. 
In the following sections, we aim to shed light on this mechanism. From Fig.~\ref{fgr:0}(a) (stretched snapshots), it is evident that damage in DNs is far more spatially delocalized than in single networks. In this section, we therefore analyse in detail how the degree of damage delocalization evolves as a function of the initial isotropic pre-stretch $\lambda_0$.

In Fig.~\ref{fgr:2}(a) we compare the density profiles of SNs and DNs to illustrate the evolution of damage.
In regime (ii)a of the DNs, instead of localizing damage to produce a macroscopic fracture, the sacrificial network forms mesoscopic damage islands (Fig.~SI6), which give rise to the stress plateau observed in the stress-stretch curve in Fig.~\ref{fgr:0}(a) and in Fig.~\ref{fgr:1}(a).
At the stretch where the matrix begins to be macroscopically loaded and starts to break (see Fig.~\ref{fgr:0}(a)), the matrix experiences a stronger local load in the zones where damage islands have already developed in the sacrificial network (Fig.~SI7). As a consequence, the damage profile of the matrix closely follows that of the sacrificial network (Fig.~\ref{fgr:A4}), resulting in similar damage islands forming within the matrix. 
Ultimately, at the end of this regime, one of these islands dominates and triggers the final fracture of the material (Fig.~\ref{fgr:2}(a), lower panel).

%%%%%%%% FIGURE 3b DENSITY and DAMAGE LOC %%%%%%%%%%%%%%%%%%%%

In Fig.~\ref{fgr:2}(b), we show the evolution of the density contrast $\Delta\rho$ (top) and of the damage localization parameter $L$ (bottom). The latter is defined as an inverse participation ratio,
$L(\lambda)=\sum_{i=1}^N n_{BS}^4(x_i,\lambda)/[\sum_{i=1}^N n_{BS}^2(x_i,\lambda)]^2$,
which quantifies the spatial localization of bond scission within a stretch interval $[\lambda,\lambda+\Delta\lambda]$, where $n_{BS}(x_i,\lambda)$ is the number of broken strands in a region of size $\Delta x=20$ and $N$ the total number of regions. Fully localized damage yields $L\to1$, whereas spatially uniform damage gives $L\to1/N$.

Initial damage occurs in a few sparse regions and progressively spreads in the entire sample as the stretch increase, thus leading to a progressive delocalization ($dL/d\lambda < 0$) in regimes (i) and (ii) until it reaches a minimum (($dL/d\lambda = 0$, indicated by the arrows in the bottom panel of Fig.~\ref{fgr:2}(b)) and eventually starts to increase ($dL/d\lambda > 0$) as a macroscopic fracture grows in the sample and induces strong density heterogeneities (regime (iii)).
The extension of the plateau observed in both $\Delta\rho$ and $L$ increases systematically with $\lambda_0$. 

From this dataset, we extract the stretch interval $\Delta \lambda ^{(ii)}$ corresponding to regime~(ii) and display it together with the stretch interval between the end of regime~(i) and the position of the maximum in the stress-stretch curve (indicated by the arrows in Fig.~\ref{fgr:1}(a)) to highlight how the extent of this regime grows with increasing initial pre-stretch. 
In analogy with the toughness measurements, the stretch range associated with the second regime increases up to a pre-stretch of $\lambda_0 = 1.8$ and then saturates, indicating this value as an optimal pre-stretch for the protocol. This interpretation is consistent with the toughness data shown in Fig.~\ref{fgr:1}(a).

%%%%%%%%%%%%%%%%%%%%%%%%%%%%%%%%%%%%%%%%%%%%%%%%%%%%%%%%%%%%%%%%%%%
%%%%%%%%%%%%%%%%%%%%%%%%%%%%%%%%%%%%%%%%%%%%%%%%%%%%%%%%%%%%%%%%%%%
%%%%%%%%%%%%%%%% RESULTS PART 4 %%%%%%%%%%%%%%%%%%%%%%%%%%%%%%%%%%%
\section{Origin of nano-voids in the sacrificial network}\label{sec:Results-4}
%%%%%%%%%%%%%%%%%%%%%%%%%%%%%%%%%%%%%%%%%%%%%%%%%%%%%%%%%%%%%%%%%%%
%%%%%%%%%%%%%%%%%%%%%%%%%%%%%%%%%%%%%%%%%%%%%%%%%%%%%%%%%%%%%%%%%%%

The persistence of a delocalized damage landscape in double networks points to a microscopic stabilization mechanism absent in single networks. In this section, we show that this stabilization arises from matrix-mediated screening of stress redistribution following individual bond-breaking events. 

To elucidate how this screening gives rise to distinct localization dynamics, we compare the response of the sacrificial network in a double network to that of a swollen single sacrificial network, focusing on the contrast between persistent damage islands and rapid localization into a single fracture region.

\begin{figure*}[t]
  \includegraphics[width=\textwidth]{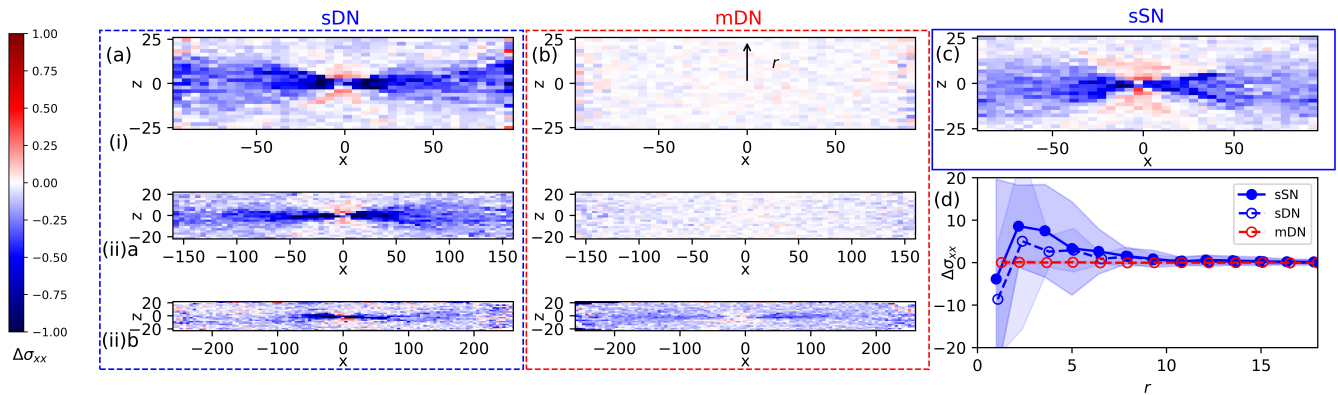}
  \caption{
  (a) Average change in stress ($xx$ component) in the sacrificial network in response to a single bond breaking event occurring in the sacrificial network for $\lambda_0=2.00$, in regimes (i) (top panel), (ii)a (middle panel) and (ii)b (bottom panel).
  (b) Response of the matrix network to bond scission occurring in the sacrificial network in the same regimes as (a).
  (c) Response for a swollen single network in regime (i), for $\lambda_0=2.00$.
  (d) Average stress change as a function of the distance $r$ to the bond breaking event in the equatorial $(yOz)$ plane (as indicated on the bottom panel of (a)), averaged over a thickness $h = 1$ along the stretching (axial) direction around the bond breaking event location for the swollen single network (solid blue), the sacrificial network inside the DN (dashed blue) and the matrix network inside the DN (dashed red).}
  \label{fgr:3}
\end{figure*}

%Response to individual bond breaking events in regime (i), (ii) and (iii)}
We measure the average stress response $\Delta \sigma_{xx}$ upon a single bond scission event in the sacrificial network, both
in the sacrificial (Fig.~\ref{fgr:3}(a)) and in the matrix (Fig.~\ref{fgr:3}(b)) networks inside a DN (sDN and mDN, respectively).
Details on the simulation and averaging protocol are provided in Appendix~\ref{app:single-bb}.

% sacrificial DN
The stress change of the sacrificial network upon bond breaking exhibits loading (positive stress change) in the equatorial plane ($yOz$) (perpendicular to the stretching direction) and relaxation (negative stress change) along the stretching direction $x$ in regimes (i), (ii)a and (ii)b (Fig.~\ref{fgr:3}(a)).
This dipolar-like response is likely to induce correlations in bond breaking events in the equatorial plane, thus favoring crack propagation perpendicular to the stretching direction.

% matrix DN
The matrix, however, does not exhibit any significant stress change upon bond breaking in the sacrificial network in regimes (i) and (ii)a (Fig.~\ref{fgr:3}(b)). This is expected since it only starts to carry the load at the end of regime~(ii)a (Fig.~\ref{fgr:0}(a)).
In regime (ii)b, however, a stress response of the matrix upon bond breaking in the sacrificial network (mainly relaxation) is observed (Fig.~\ref{fgr:3}(b), bottom panel), providing a direct evidence of load sharing upon damage at large stretch in double networks, as hypothesized by Tauber et al. \cite{tauber2021sharing}. 

% comparison with sacrificial SN
To further investigate the role played by the matrix at small stretch (regime (i)), we compare the response to bond scission of the sacrificial network inside the DN (sDN) to the response of a single sacrificial network (sSN) (Fig.~\ref{fgr:3}(c)).
Fig.~\ref{fgr:3}(d) depicts the average stress change $\Delta \sigma_{xx}$ as a function of the distance to the bond breaking event in the radial direction $r$, where the loading ($\Delta \sigma_{xx}>0$) is maximal.
%(perpendicular to the stretching direction $x$, in the equatorial plane ($yOz$), as indicated in the top panel of Fig.~\ref{fgr:3}(b)).
Interestingly, $\Delta \sigma_{xx}$  is larger in the single sacrificial network (sSN) compared to the double network (sDN), suggesting that the presence of the matrix induces a screening of the sacrificial network response to scission events even before it starts to be (macroscopically) loaded.
These findings point towards a protective role of the matrix at small stretch in DNs, 
lowering the load on the sacrificial network neighbouring strands upon bond scission compared to single sacrificial networks, and thus limiting correlated bond scission in sDN.

We propose to further investigate how this screening mechanism leads to delocalized damage at low stretch by studying the evolution of an isolated damage island is DN and sSN.

%%%%%%%%%%%%%%%%%%%%%%%%%%%%%%%%%%%%%%%%%%%%%%%%%%%%%%%%%%%%%%%%%%%%%%%%%%%%%%%%%%%%%%%%%%%%%%%%%%%%%%%%%%%%
%%%%%%%%%%%%%%%%%%%%%%%%%%%%%%  FIGURE 4 %%%%%%%%%%%%%%%%%%%%%%%%%%%%%%%%%%%%%%%%%%%%%%%%%%%%%%%%%%%%%%%%%%%
%%%%%%%%%%%%%%%%%%%%%%%%%%%%%%%%%%%%%%%%%%%%%%%%%%%%%%%%%%%%%%%%%%%%%%%%%%%%%%%%%%%%%%%%%%%%%%%%%%%%%%%%%%%%
%%%%%%%%%%%%%%%%%%%%%%%%%%%%%%%%%%%%%%%%%%%%%%%%%%%%%%%%%%%%%%%%%%%%%%%%%%%%%%%%%%%%%%%%%%%%%%%%%%%%%%%%%%%%
%%%%%%%%%%%%%%%%%%%%%%%%%%%%%%%%%%%%%%%%%%%%%%%%%%%%%%%%%%%%%%%%%%%%%%%%%%%%%%%%%%%%%%%%%%%%%%%%%%%%%%%%%%%%
%%%%%%%%%%%%%%%%%%%%%%%%%%%%%%%%%%%%%%%%%%%%%%%%%%%%%%%%%%%%%%%%%%%%%%%%%%%%%%%%%%%%%%%%%%%%%%%%%%%%%%%%%%%%

%%%%%%%%%%%%%%%%%%%%%%%%%%%%%%%%%%%%%%%%%%%%%%%%%%%%%%%%%%%%%%%%%%%%%%%%%%%%%%%%%%%%%%%%%%%%%%%%%%%%%%%%%%%%
%%%%%%%%%%%%%%%%%%%RESULTS PART 5 %%%%%%%%%%%%%%%%%%%%%%%%%%%%%%%%%%%%%%%%%%%%%%%%%%%%%%%%%%%%%%%%%%%%%%%%%%
%%%%%%%%%%%%%%%%%%%%%%%%%%%%%%%%%%%%%%%%%%%%%%%%%%%%%%%%%%%%%%%%%%%%%%%%%%%%%%%%%%%%%%%%%%%%%%%%%%%%%%%%%%%%
\section{Evolution of an isolated nano-void under stretch}\label{sec:Results-5}
%%%%%%%%%%%%%%%%%%%%%%%%%%%%%%%%%%%%%%%%%%%%%%%%%%%%%%%%%%%%%%%%%%%%%%%%%%%%%%%%%%%%%%%%%%%%%%%%%%%%%%%%%%%%
%%%%%%%%%%%%%%%%%%%%%%%%%%%%%%%%%%%%%%%%%%%%%%%%%%%%%%%%%%%%%%%%%%%%%%%%%%%%%%%%%%%%%%%%%%%%%%%%%%%%%%%%%%%%

\begin{figure*}[t]
  \includegraphics[width=0.8\textwidth]{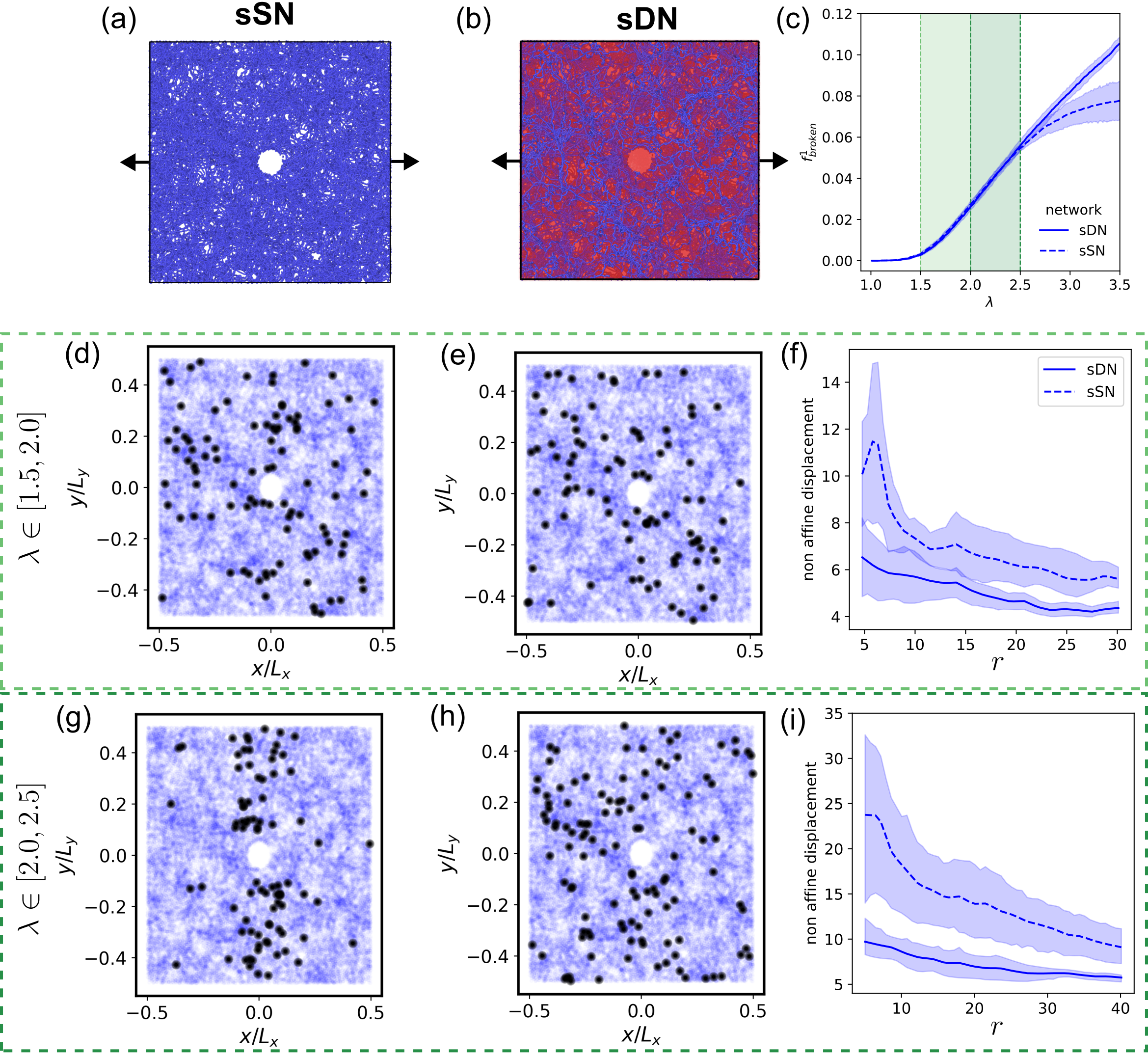}
  \caption{Shown are snapshots of typical initial configurations with a circular micro-crack in the (a) swollen single network and in the (b) sacrificial network of the DN . In (c) we show the comparative curves of the fraction of broken bonds in double and single swollen network as a function of stretch, both have been isotropically pre-stretched with $\lambda_0=2.0$. In (d) and (g) we see the swollen single network together with the positions of the broken bonds accumulated in the strain region shaded respectively in light-gray and gray in (c). (e) and (h) display the same data for the sacrificial network of the DN. (f) and (i) show the absolute non-affine displacements as a function of the radial distance r from the center of the micro-crack, respectively to the two shaded stretch regimes in (c).}
  \label{fgr:4}
\end{figure*}

To investigate how an isolated damage island evolves in the double network, we perform idealised 3d simulations in which a cylindrical region of radius $R$ (with its cylindrical axis parallel to $z$, perpendicular to the stretch direction) is damaged within the sacrificial network. This is achieved by cutting all bonds inside this region and manually removing the corresponding monomers before carrying out the stretching protocol. To isolate the influence of the matrix on the dynamics of the damaged zone, we compare this DN configuration to its swollen single network counterpart (sSN). Typical initial configurations are shown in Fig.~\ref{fgr:4}(a) and (b).

In Fig.~\ref{fgr:4}(c) we show that the fraction of strand scission in the sacrificial network inside the DN (sDN) and in the sSN is essentially identical for the two initial configurations up to a stretch of about $\lambda = 2.5$ (where it reaches $\sim 5 \%$), suggesting that the system responds in a very similar manner to the applied deformation (see section~\ref{sec:Results-1}).
This stretch range corresponds to regime~(i), where the macroscopic stress is still dominated by the sacrificial network and the matrix remains effectively unloaded (Fig.~\ref{fgr:0}(b) and Fig.~SI11). As we will demonstrate in the following, however, this apparent similarity masks fundamentally different underlying mechanisms.

A detailed spatial analysis of the accumulated bond breaking events in the stretch windows $\lambda \in [1.5, 2.0]$ and $\lambda \in [2.0, 2.5]$ reveals that, at low stretch ($\lambda \in [1.5, 2.0]$), bond breaking events are homogeneously distributed throughout the system in both DNs and sSNs. This indicates that, in this regime, the initial preparation of the networks determines the location of the first bond breaking events (consistent with Fig.~\ref{fgr:A3}).
In contrast, for $\lambda \in [2.0, 2.5]$, a clear distinction emerges between the DN and the sSN. The latter exhibits a strongly localized damage stripe along the $y$ direction, perpendicular to both the imposed stretch and the cylindrical axis of the initially damaged zone. In the double network, however, the accumulated bond breaking events remain broadly distributed: the matrix effectively suppresses the localisation that is otherwise observed in the single network.  

The origin of this difference lies in the screened response of the surroundings to bond breaking events, as already evidenced in the bulk dynamics discussed previously.
% Non Affine Displacement
When analysing the non-affine displacements during the two stretch windows, we find that even in the earliest stretching regime the response of the sacrificial network to bond breaking is significantly reduced compared to its single network counterpart. At small stretch, this difference does not visibly influence the overall dynamics, but as the stretch increases the contrast becomes amplified. 
In the swollen single network the elastic response facilitates the onset of localisation, whereas in the double network, the matrix, although not yet macroscopically loaded, strongly damps the response of the sacrificial network to bond breaking events and thereby prevents localisation. The matrix effectively converts an unstable elastic defect into a marginally stable one.

%%%%%%%%%%%%%%%%%%%%%%%%%%%%%%%%%%%%%%%%%%%%%%%%%%%%%%%%%%%%%%%%%%%%%%%%%%%%%%%%%%%%%%%%%%%%%%%%%%%%%%%%%%%%
%%%%%%%%%%%%%%%%%%%%%%%%%%%%%%  FIGURE 5 %%%%%%%%%%%%%%%%%%%%%%%%%%%%%%%%%%%%%%%%%%%%%%%%%%%%%%%%%%%%%%%%%%%
%%%%%%%%%%%%%%%%%%%%%%%%%%%%%%%%%%%%%%%%%%%%%%%%%%%%%%%%%%%%%%%%%%%%%%%%%%%%%%%%%%%%%%%%%%%%%%%%%%%%%%%%%%%%
%%%%%%%%%%%%%%%%%%%%%%%%%%%%%%%%%%%%%%%%%%%%%%%%%%%%%%%%%%%%%%%%%%%%%%%%%%%%%%%%%%%%%%%%%%%%%%%%%%%%%%%%%%%%
%%%%%%%%%%%%%%%%%%%%%%%%%%%%%%%%%%%%%%%%%%%%%%%%%%%%%%%%%%%%%%%%%%%%%%%%%%%%%%%%%%%%%%%%%%%%%%%%%%%%%%%%%%%%
%%%%%%%%%%%%%%%%%%%%%%%%%%%%%%%%%%%%%%%%%%%%%%%%%%%%%%%%%%%%%%%%%%%%%%%%%%%%%%%%%%%%%%%%%%%%%%%%%%%%%%%%%%%%

\section{Discussion and conclusions}

Our simulations reveal a sequence of deformation regimes that together provide a unified picture of damage evolution in double networks. While the early response is dominated by the pre-stretched sacrificial network and closely resembles that of swollen single networks, a crucial qualitative difference emerges: local stress perturbations induced by sacrificial bond scission are partially screened by the matrix. This screening suppresses the anisotropic stress redistribution characteristic of single networks, preventing rapid defect growth. As a result, damage accumulates in the form of multiple stabilized micro-damage zones. At larger stretch, when the matrix becomes load-bearing, these pre-existing zones bias matrix failure in a spatially correlated yet delocalized manner, enabling efficient energy dissipation prior to final fracture.

The exceptional toughness of double-network polymers is commonly attributed to the interplay between sacrificial bond scission and load transfer to a more extensible matrix network. Early mechanoluminescence experiments by Millereau \emph{et al.}~\cite{Millereau2018} demonstrated that fracture proceeds through two distinct stages, with an initial regime of distributed sacrificial bond breaking followed by more localized damage associated with large energy dissipation. While these experiments established the existence of a two-stage fracture scenario, the spatial organization of damage and the role of early scission events in shaping subsequent localization remained unclear.

Subsequent experimental and numerical studies by the groups of van der Gucht and Cipelletti~\cite{tauber2022stretchy,orr2025probing} revealed that double networks exhibit early-onset, spatially delocalized microscopic rearrangements extending far ahead of macroscopic failure. These results highlighted the role of efficient stress redistribution and enhanced microscopic dynamics in delaying crack propagation.
However, damage in these studies was primarily characterized through dynamical activity and density fluctuations, leaving the causal connection between sacrificial bond scission, its spatial organization, and the onset of matrix failure implicit.

From a theoretical standpoint, Walker and Fielding~\cite{walker2025toughness} proposed a minimal mesoscale model in which strong inter-network coupling reduces the amplitude of an Eshelby-like stress redistribution, suppressing crack-like avalanches and promoting diffuse damage. While this framework captures an important limiting mechanism for inhibiting localization, it does not resolve the dynamical emergence of damage organization in realistic three-dimensional polymer networks, nor the transition to matrix-dominated failure.

Our results build on and extend these works by resolving the full spatiotemporal evolution of damage throughout deformation. We show that early sacrificial bond scission is not random, but instead generates a dynamically delocalized damage landscape through matrix-mediated screening of local stress perturbations. This delocalization stabilizes multiple damage islands and actively prepares the system for a subsequent regime in which matrix bonds fail in a spatially correlated yet delocalized manner. By explicitly linking early sacrificial damage, stress redistribution, and later matrix failure, our work provides a unified microscopic mechanism that connects experimental observations~\cite{Millereau2018,tauber2022stretchy,orr2025probing} and mesoscale theories based on stress screening~\cite{walker2025toughness}.

%%%%%%%%%%%%% PERSPECTIVES %%%%%%%%
Our findings open a new avenue of research in the field. Experimentally, matrix-mediated screening could be probed by measuring spatial correlations of mechanophore activation following localized bond scission, or by comparing displacements around controlled defects in single and double networks using dynamic light scattering.

An important extension of the present work is to study hierarchical interpenetrating networks beyond the double-network architecture, such as triple- and higher-order network elastomers. A key open question is whether each additional network introduces a new stage of stress screening and load sharing, and how this hierarchy controls damage delocalization, the onset of localization, and the resulting toughness. Addressing these questions might lead into an optimizing strategy of the mechanical performances of multi-networks, but it requires systematically linking network architecture, pre-stretch, and stiffness contrast to the spatial organization of damage and energy dissipation.

A further extension of this framework is to address fracture dynamics in hydrogel networks, where solvent content and poroelastic effects introduce additional dissipation channels. Building on the microscopic mechanisms identified here, this requires elucidating how stress screening, bond scission, and load redistribution couple to solvent-mediated energy transport near a propagating crack. Understanding how these processes control crack initiation, intermittency, and velocity selection remains an open challenge \cite{zheng2021chain}. 

Finally, there is a need into developing continuum-scale models that capture the macroscopic consequences of the microscopic mechanisms identified here. In particular, constitutive descriptions should incorporate matrix-mediated screening of local stress perturbations, the emergence of spatially heterogeneous damage landscapes, and the transition to macroscopic load sharing biased by pre-existing damage. Embedding these ingredients into finite-element frameworks would enable quantitative comparisons with experiments and provide predictive tools for designing tough, damage-tolerant polymer networks.

\begin{acknowledgments}
We acknowledge valuable discussions with Nicholas Orr, Luca Cipelletti, Laurence Ramos, Costantino Creton, Ana\"el Lema\^\i tre, Jasper Van der Gucht, Sebastian Pfaller, Maximilian Ries, José Ruiz Franco Martin Lenz, Franck Vernerey, Emanuela Del Gado and Peter Olmsted. 
This work was supported by the French ANR, grant n.~ANR-20-CE06-0028 (MultiNet project). M.B.~acknowledges support by the French ANR grant ANR-22-CE06-0036-03 (Microfat project). L. O. acknowledges support from the European Union's Horizon Europe research and innovation programme through the Marie Sklodowska-Curie grant agreement No. 101204391.
J.T.~acknowledges the financial support by the China Scholarship Council (CSC) (No. 201904890006) and by the National Natural Science Foundation of China (No. 52303029).  Numerical simulations have been performed on the GRICAD infrastructure (https://gricad.univ-grenoble-alpes.fr), which is supported by the Grenoble research communities. 
\end{acknowledgments}

%%%%%%%%%%%%%%%%%%%%%%%%%%%%%%%%%%%%%%%%%%%%%%%%%%%%%%%%%%%%%%%%%%%%%%%%%%%%%%%%%%%%%%%%%%%%%%%%%%%%%%%%%%%%%%%%%%%%%%%%
%%%%%%%%%%%%%%%%%%%%%%%%%%%%%%%%%%%%%%%%%%%%%%%%%%%%%%%%%%%%%%%%%%%%%%%%%%%%%%%%%%%%%%%%%%%%%%%%%%%%%%%%%%%%%%%%%%%%%%%%
\appendix %%%%%%%%%%%%%%%%%%%%%%%%%%%%%%%%%%%%%%%%%%%%%%%%%%%%%%%%%%%%%%%%%%%%%%%%%%%%%%%%%%%%%%%%%%%%%%%%%%%%%%%%%%%%%%
%%%%%%%%%%%%%%%%%%%%%%%%%%%%%%%%%%%%%%%%%%%%%%%%%%%%%%%%%%%%%%%%%%%%%%%%%%%%%%%%%%%%%%%%%%%%%%%%%%%%%%%%%%%%%%%%%%%%%%%%
%%%%%%%%%%%%%%%%%%%%%%%%%%%%%%%%%%%%%%%%%%%%%%%%%%%%%%%%%%%%%%%%%%%%%%%%%%%%%%%%%%%%%%%%%%%%%%%%%%%%%%%%%%%%%%%%%%%%%%%%
\setcounter{figure}{0} % Réinitialiser le compteur
\renewcommand{\thefigure}{A\arabic{figure}} % Changer le format de numérotation

%%%%%%%%%%%%%%%%%%%%%%%%%%%%%%%%%%%%%%%%%%%%%%%%%%%%%%%%%%%%%%%%%%%%%%%%%%%%%%%%%%%%%%%%%%%%%%%%%%%%%%%%%%%%%%%%%%%%%%%%
\section{Model details}\label{app:model-details} %%%%%%%%%%%%%%%%%%%%%%%%%%%%%%%%%%%%%%%%%%%%%%%%%%%%%%%%%%%%%%%%%%%%%%% %%%%%%%%%%%%%%%%%%%%%%%%%%%%%%%%%%%%%%%%%%%%%%%%%%%%%%%%%%%%%%%%%%%%%%%%%%%%%%%%%%%%%%%%%%%%%%%%%%%%%%%%%%%%%%%%%%%%%%%%

\subsection{Interaction potentials}

The interaction between beads separated by a distance $r$ is described by the Weeks--Chandler--Andersen (WCA) potential, a truncated and shifted Lennard-Jones potential:%\cite{Grest1996}
\begin{equation}
\label{eq_imp:LJ}
U_{\mathrm{WCA}}(r) =
\begin{cases}
4\varepsilon \left[ \left(\frac{\sigma}{r}\right)^{12}
- \left(\frac{\sigma}{r}\right)^{6} \right] + \delta, & r \le r_c, \\[6pt]
0, & r > r_c .
\end{cases}
\end{equation}
All length- and distance-related quantities are expressed in units of the bead radius~$\sigma$.

To allow bond scission when the bond length exceeds $R_0$ during uniaxial extension, the bond potential is replaced by a quartic potential of the form
\begin{eqnarray}
U_\mathrm{bond}(r) = K (r-R_c)^3(r-R_c - B) + U_0 + U_\mathrm{WCA}(r)  
\label{bond_break}
\end{eqnarray}
where $K = 2351$, $R_c = R_0$, $U_0 = 92.74467$, $B = -0.7425$, and $\delta = \epsilon$ in the WCA potential. This parametrization prevents chain crossing and reproduces the same equilibrium bond length as the FENE potential \cite{ge2013molecular}.

The Kremer-Grest model further allows tuning of chain stiffness to reproduce the rheological properties of real polymers \cite{everaers2020kremer}. In our simulations, bonded particles additionally interact via a three-body angular potential to capture bending rigidity:
\begin{eqnarray}
U_\mathrm{angular}(\theta) = K_\theta \left ( 1 + \cos{(\theta)} \right )  
\label{eq-angular-potential}
\end{eqnarray}
where $\theta$ is the angle formed by three consecutive monomers along the polymer chain, and $K_\theta=1.276$ is chosen to mimic the bending rigidity of poly(ethyl acrylate) (PEA) multiple elastomer networks \cite{Millereau2018,everaers2020kremer}.

%%%%%%%%%%%%%%%%%%%%%%%%%%%%%%%%%%%%%%%%%%%%%%%%%%%%%%%%%%%%%%%%%%%%%%%%%%%%%%%%%%%%%%%%%%%%%%%%%%%%%%%%%%%%%%%%%%%%%%%%
\subsection{Synthesis protocol} %%%%%%%%%%%%%%%%%%%%%%%%%%%%%%%%%%%%%%%%%%%%%%%%%%%%%%%%%%%%%%%%%%%%%%% %%%%%%%%%%%%%%%%%%%%%%%%%%%%%%%%%%%%%%%%%%%%%%%%%%%%%%%%%%%%%%%%%%%%%%%%%%%%%%%%%%%%%%%%%%%%%%%%%%%%%%%%%%%%%%%%%%%%%%%%

%%%%%%%%%%%%%%%%%%%%%%%%%%%%%%%%%%%FIGURE A1 : synthesis, strand length and strand stretch %%%%%%%%%%%%%%%%%%%%%%%%%%

\begin{figure}[t]
  \includegraphics[width=\columnwidth]{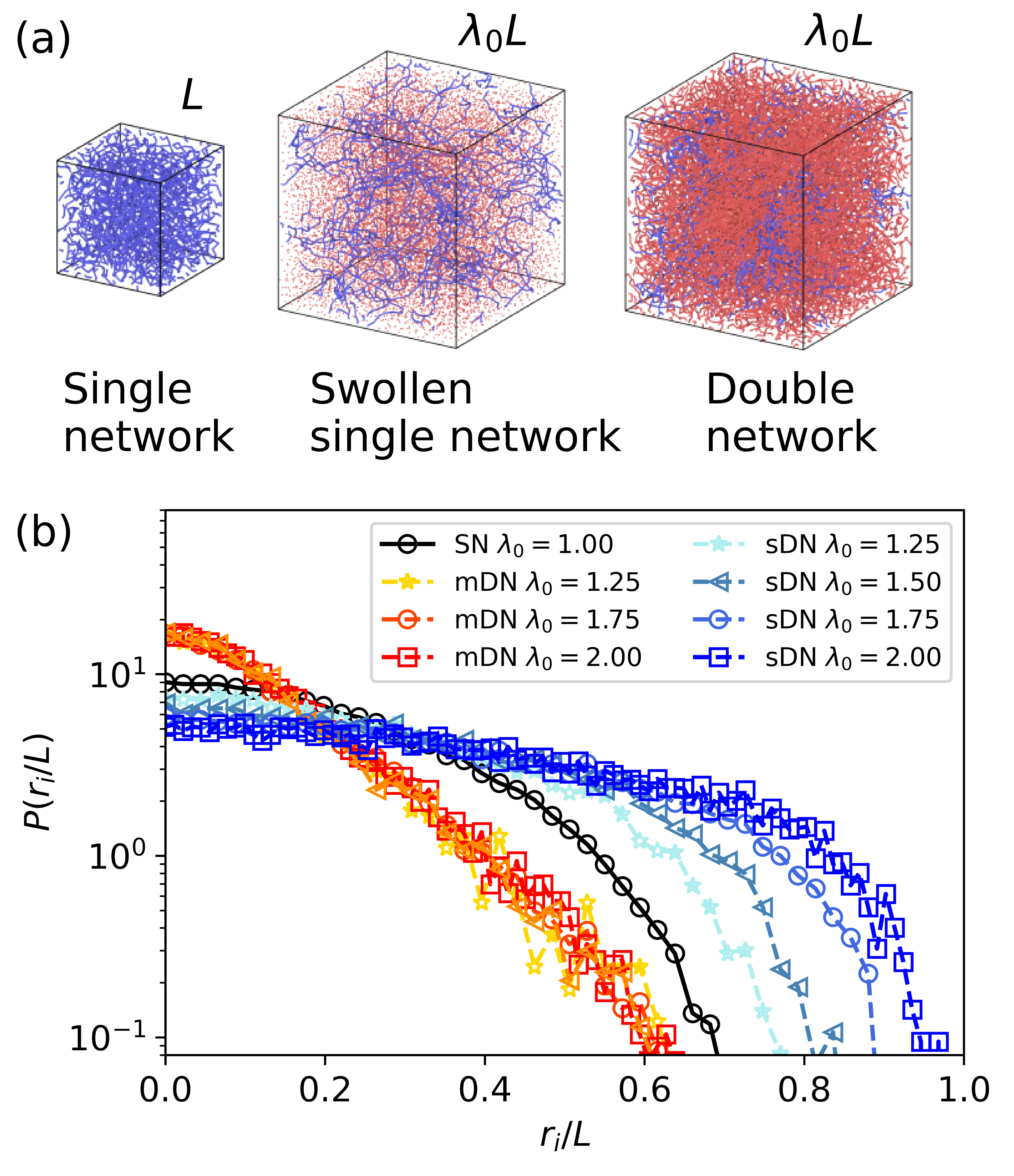}
  \caption{ (a) Synthesis protocol : a single network of size  of size $L \times L \times L$ is first synthesized (left) and then swollen using the monomers of a second network with a swelling ratio $\lambda_0$ (middle) before the second network is polymerized and cross-linked (right).
  (b) Initial strand stretch (obtained from thermally averaged configurations) in the sacrificial network sDN (black and blue symbols) and in the matrix network mDN (yellow, orange and red symbols) inside the DN (mDN) for various values of $\lambda_0$.}
  \label{fgr:A1}
\end{figure}
%%%%%%%%%%%%%%%%%%%%%%%%%%%%%%%%%%%%%%%%%%%%%%%%%%%%%%%%%%%%%%%%%%%%%%%%%%%%%%%%%%%%%%%%%%%%%%%%%%%%%%%%%%%%%%%%%%%%%

Following experimental synthesis protocols for elastomer networks \cite{Millereau2018}, we developed a computational procedure to construct disordered polymer networks, both single and double, using a two-step radical-like polymerization scheme \cite{perez2008polymer} in which the bonds are progressively created as a radical monomer approaches another monomer. The procedure start with a mixture of dimer crosslinker molecules (made of two bonded particles) with concentration $c_1$ and $c_2$ for the first and the second network respectevely. As illustrated in Fig.~\ref{fgr:A1}(a), the first network is generated by adding $N_1^{\mathrm{monomer}}$ monomers and radical beads at concentration $c_1^r$, such that $N_1^{\mathrm{radical}} = c_1^r N_1^{\mathrm{monomer}}$. The overall monomer density is $\rho_0^1 = 0.8$. Radical-like polymerization and crosslinking are then performed, creating FENE bonds between bonded monomers, which yields a disordered single network of size $L \times L \times L$. This network will be referred to as the sacrificial network throughout the manuscript.

A double network is created by swelling this first network through the addition of $N_2^{\mathrm{monomer}}$ monomers and $N_2^{\mathrm{radical}} = c_2^r N_2^{\mathrm{monomer}}$ radical beads, together with crosslinker molecules. To avoid excessive overlaps upon insertion, the repulsive interaction between the second-network monomers is initially set to a soft cosine potential $E(r) = A \left( 1 + \cos \pi \frac{r}{r_c} \right)$, with prefactor $A$ progressively increased in the $NPT$ ensemble (at $P=3.8$). The potential is then switched to the WCA form, and the system is relaxed in the $NVT$ ensemble at an expanded box size of $\lambda_0 L \times \lambda_0 L \times \lambda_0 L$, corresponding to the swelling ratio $\lambda_0$ (see Fig.~\ref{fgr:A1}(a), middle panel). A second radical-like polymerization step, with simultaneous crosslinking, is then performed to form the matrix network within the already swollen first (now sacrificial) network, resulting in an interpenetrated double-network structure (Fig.~\ref{fgr:A1}(a), right panel). Finally, FENE bonds are replaced by quartic bonds to enable bond scission, and the system is equilibrated with $R_0 = 1.5 \, \sigma$.

%\subsection{Networks structure}
All networks (single, sacrificial, and matrix) display an exponential distribution of strand lengths, consistent with random crosslink placement and Flory–Stockmayer theory \cite{flory1953principles,rubinstein2003polymer}
% CITE SI1
(see Supplementary Information, Fig.~SI1).

% System size
The total number of particles in the system is $N = 4.4 \times 10^5$, with system size $\lambda_0 L = 82 $ (in units of bead diameter $\sigma$, which is not mentioned otherwise). 
% Crosslink concentration
We use crosslink concentrations of $c_1 = 5\%$ and $c_2 = 2\%$, together with radical concentrations of $c_1^r = 0.5\%$ and $c_2^r = 0.1\%$, yielding average strand lengths of $\langle N_1 \rangle \simeq 10$ monomers for the first network and $\langle N_2 \rangle \simeq 50$ monomers for the second network.

The swelling process that precedes the formation of the second network introduces an isotropic pre-stretch % Magali : included Jiting comment: the strands don't stretch affinely during swelling !
of the first (sacrificial) network. This is quantified in Fig.~\ref{fgr:A1}(b), which shows the distribution of the initial strand stretch, defined as $r_i/L$, averaged over all spatial directions. Here $L=(n-1)b$ is the strand contour length, with $n$ the number of monomers per strand and $b\simeq 0.96$ the equilibrium bond length. As $\lambda_0$ increases, a growing fraction of first-network strands approach their contour length, while the second (matrix) network, synthesized after swelling, shows a stretch distribution centered around its equilibrium value. 
In addition, the swelling step induces the formation of nanoscale voids within the sacrificial network, which are subsequently filled by the matrix during polymerization. This structural feature was independently confirmed by dynamic light scattering measurements on double networks\cite{na2004structural}. See Figure \ref{fgr:0}(a) for typical initial configurations of the single network, the double network and the swollen single network (identical to the sacrificial network in the double network).

%%%%%%%%%%%%%%%%%%%%%%%%%%%%%%%%%%%%%%%%%%%%%%%%%%%%%%%%%%%%%%%%%%%%%%%%%%%%%%%%%%%%%%%%%%%%%%%%%%%%%%%%%%%%%%%%%%%%%%%%
\section{Uniaxial extension simulations}\label{app:stretch-protocol} %%%%%%%%%%%%%%%%%%%%%%%%%%%%%%%%%%%%%%%%%%%%%%%%%%% %%%%%%%%%%%%%%%%%%%%%%%%%%%%%%%%%%%%%%%%%%%%%%%%%%%%%%%%%%%%%%%%%%%%%%%%%%%%%%%%%%%%%%%%%%%%%%%%%%%%%%%%%%%%%%%%%%%%%%%%

Uniaxial stretching of the simulation box is carried out using a stepwise protocol in which the box dimensions and particle positions are affinely rescaled by $\Delta \lambda = 1\%$ along the $x$-axis at each step. 
After each deformation step, the system is relaxed, corresponding to a strain rate of $\dot \lambda = 4 \cdot 10 ^{-5} \tau ^{-1}$ under Langevin dynamics at $T=1.0$, with damping coefficient $\xi = 1.0$. A Berendsen barostat is applied independently in the $y$ and $z$ directions to maintain a constant pressure of $P = 3.5$, corresponding to the equilibrium pressure prior to deformation. 
Periodic boundary conditions are applied in all directions. The integration time step is $\Delta t = 0.005$. See Fig.\ref{fgr:0}(a) for typical stretched configurations ($\lambda=9$) for single, double and sacrificial networks.

The stress response is computed from the virial stress, and the true stress is given by  $\sigma_{T} = \sigma_{xx} - \frac{1}{2}(\sigma_{yy} + \sigma_{zz})$.
All simulations are performed using the \textsc{LAMMPS} package \cite{Lammps}, and visualization is carried out with the \textsc{OVITO} software \cite{ovito}.

% Stress-stretch curve measurements
The average macroscopic stress $\sigma_{T}$ (at a given stretch value $\lambda$) is measured by averaging over a time window $T_\mathrm{av}=150\tau$ in steady state. 
% CITE SI2
The different contributions of the interaction potential to the stress-strain curve are depicted in Fig.~SI2 and one can see that the macroscopic stress response is dominated by the bond interactions.

\section{Damage dynamics}

\subsection{Link between initial structure and sacrificial damage}

%%%%%%%%%%%%%%%%%%%%%%%%%%%%%%% FIGURE A2 %%%%%%%%%%%%%%%%%%%%%%%%%%%%%%%%%%%%%%%%%%%%%% 
\begin{figure}[t]
  \includegraphics[width=\columnwidth]{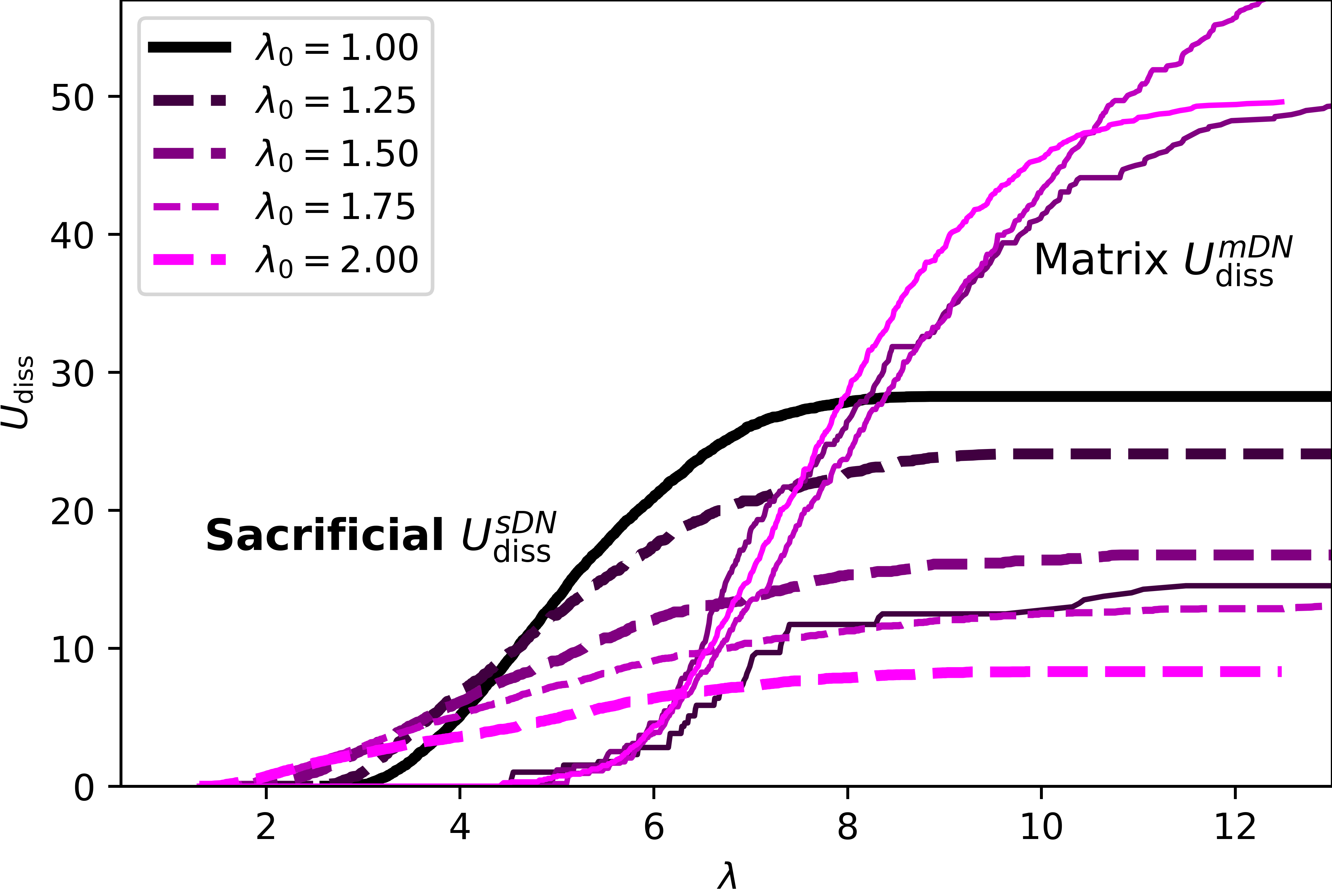}
  \caption{Energy dissipated (per unit volume) by bond breaking in the sacrificial (thick dashed line) and matrix (thin solid line) networks inside the DN, respectively using $U_\mathrm{diss}^{sDN} = u_1 n_1$ and $U_\mathrm{diss}^{mDN} = u_2 n_2$ for different values of $\lambda_0$.}
  \label{fgr:A2}
\end{figure}
%%%%%%%%%%%%%%%%%%%%%%%%%%%%%%%%%%%%%%%%%%%%%%%%%%%%%%%%%%%%%%%%%%%%%%%%%%%%%%%%%%%%%%%%%%%%%%%%

\begin{figure}[t]
  \includegraphics[width=\columnwidth]{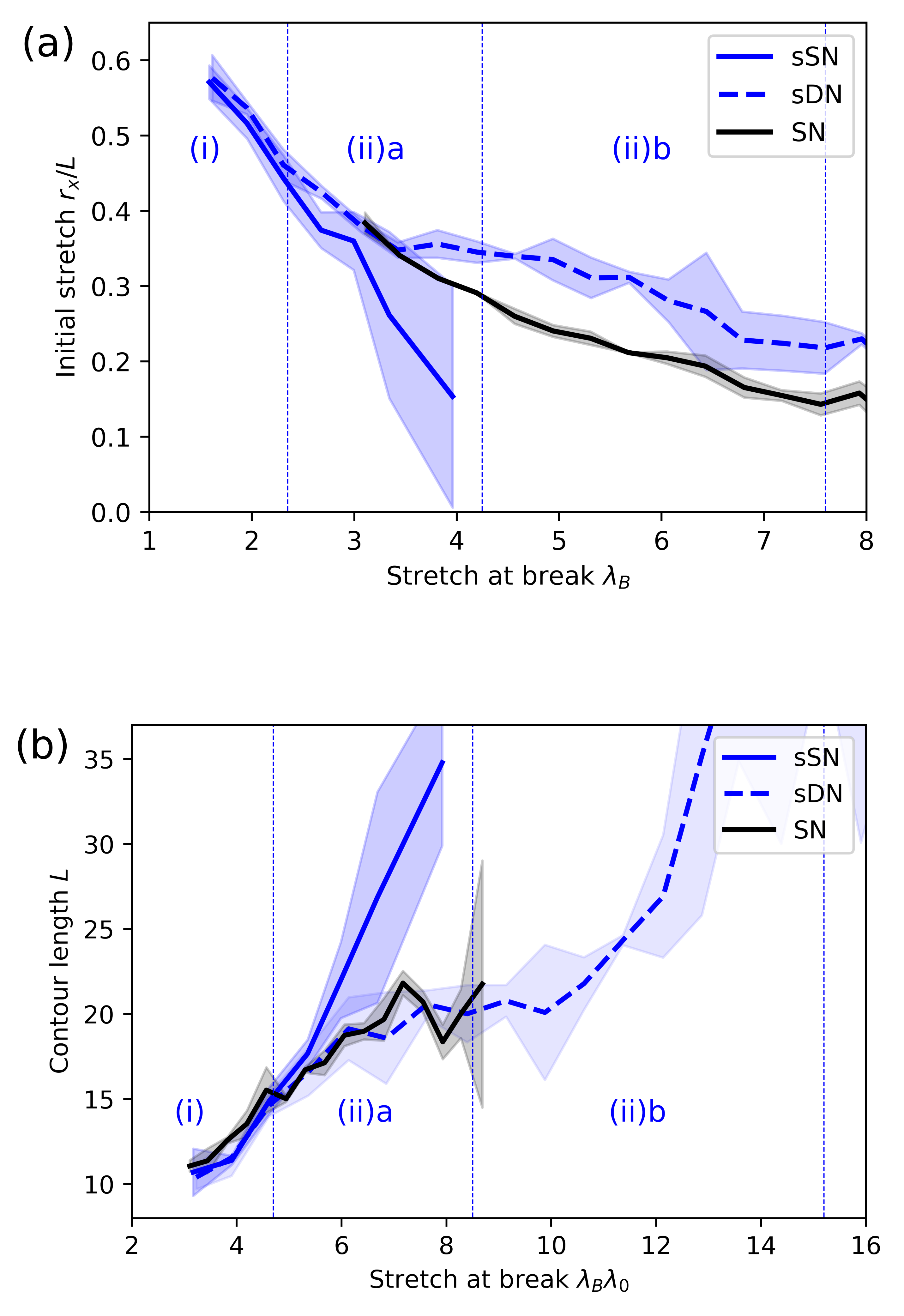}
  \caption{
  (a) Initial stretch of the broken strand as a function of the stretch at break for SN samples (black), swollen SN (sSN) (solid blue line) and sacrificial networks inside DN samples (sDN) (dashed blue lines).
  (b) Contour length of the broken strand as a function of the stretch at break for the same samples.}
  \label{fgr:A3}
\end{figure}

Fig.~\ref{fgr:A3} depicts the average initial strand stretch and strand contour length in the sacrificial network inside the DN (sDN) and in the SN as a function of the stretch at break.
This analysis is performed for each individual broken strand (of each network) and then results are binned (following Tauber et al. \cite{tauber2021sharing}).

We identify two different behaviors in this graph: in regimes (i) and (ii)a ($\lambda \lambda_0 < 8.5$), there is a strong correlation between the properties of the undeformed sacrificial network and the stretch at break.
The correlation with the initial network configuration is lost for $\lambda \lambda_0>8.5$ (end of regime (ii)a), indicating that stress redistribution mechanisms dominate in regime (ii)b.

%%%%%%%%%%%%%%%%%%%%%%%%%%%%%%% FIGURE A3 %%%%%%%%%%%%%%%%%%%%%%%%%%%%%%%%%%%%%%%%%%%%%% 
%%%%%%%%%%%%%%%%%%%%%%%%%%%%%%%%%%%%%%%%%%%%%%%%%%%%%%%%%%%%%%%%%%%%%%%%%%%%%%%%%%%%%%%%%%%%%%%%

\subsection{Density and damage localization} 

Density profiles, $\rho(x)$, along the stretching direction $x$ are computed by integrating the local density along $y$ and $z$, the directions perpendicular to the stretching direction, and then by binning along $x$ using a bin width $\Delta x = 10$.
The density localization is computed as $\Delta \rho = \rho_\mathrm{max} - \rho_\mathrm{min}$, with $\rho_\mathrm{max}$ and $\rho_\mathrm{min}$ the maximum and minimum densities in the profile $\rho(x)$, respectively. Since the average density $\rho_0 = 0.8$, a density localization $\Delta \rho  > 0.8$ means that the network has one or several regions with a density $\rho_\mathrm{min} \simeq 0$ (fractured regions), with other regions being densified after fracture, yielding $\rho_\mathrm{max} > \rho_0$

\begin{figure}[t]
  \includegraphics[width=\columnwidth]{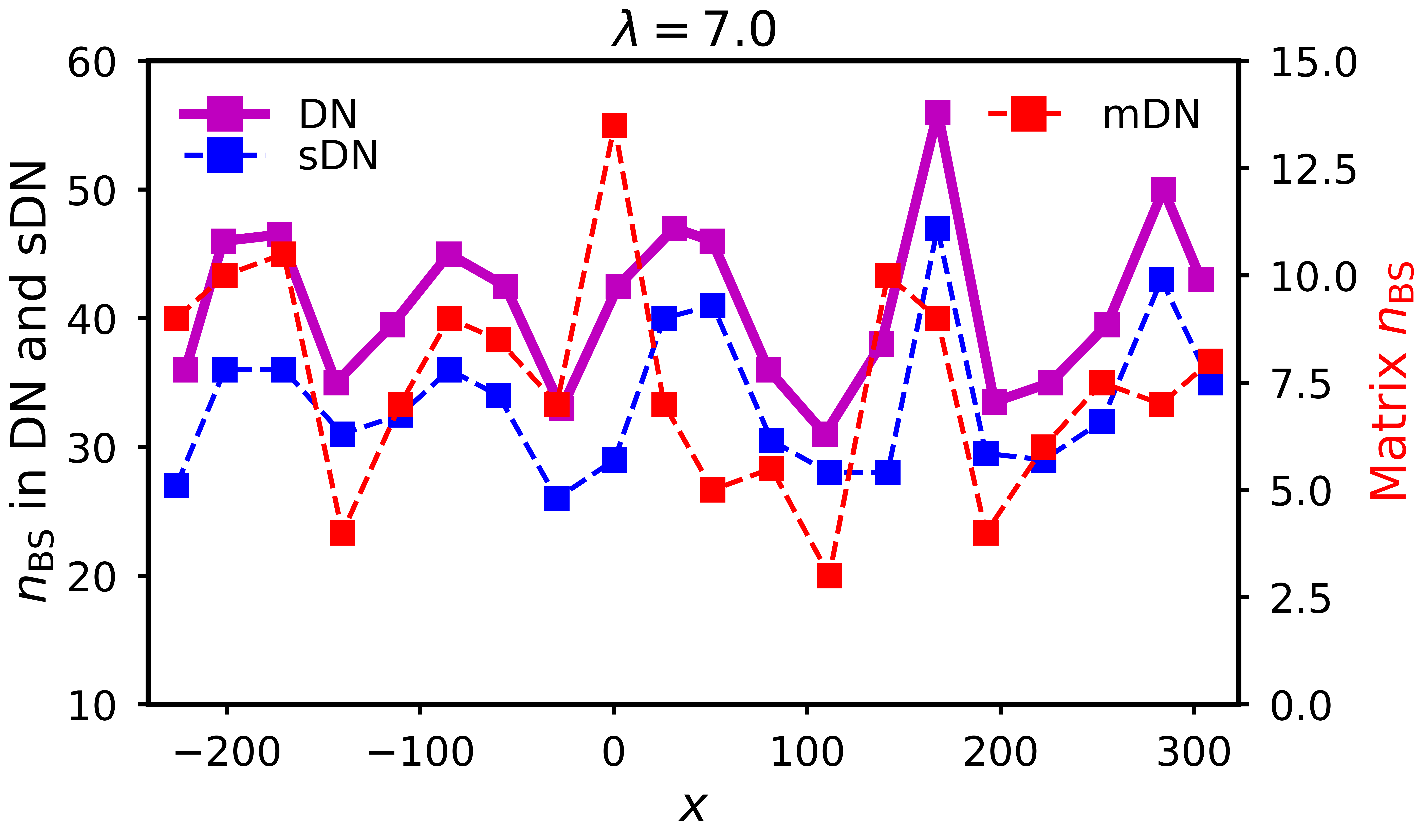}
  \caption{Profiles of the number of broken strands along the stretching direction (accumulated damage throughout the simulation) up to a stretch value $\lambda = 7.0$ in the sacrificial network (blue squared symbols) and in the matrix network (red squared symbols).}
  \label{fgr:A4}
\end{figure}

Broken bonds profiles (shown in Fig.~\ref{fgr:A4}) are computed by integrating the position of broken bonds (in the last saved configuration before scission) along $y$ and $z$, the directions perpendicular to the stretching direction, and then by binning along $x$ using a bin width $\Delta x_B = 15$ (we take larger bins for bond breaking profiles than for density profiles due to the smaller amount of broken bonds per bin).

Fig.~\ref{fgr:A4} shows examples of accumulated broken bonds profiles for the same network up to a stretch $\lambda = 7.0$ (near the end of regime (ii)b), where bond scission has occurred both in the sacrificial network and in the matrix network.

%%%%%%%%%%%%%%%%%%%%%%%%%%%%%%% FIGURE A4 %%%%%%%%%%%%%%%%%%%%%%%%%%%%%%%%%%%%%%%%%%%%%% 
%%%%%%%%%%%%%%%%%%%%%%%%%%%%%%%%%%%%%%%%%%%%%%%%%%%%%%%%%%%%%%%%%%%%%%%%%%%%%%%%%%%%%%%%%%%%%%%%

Note that the participation ratio shown in Fig.~\ref{fgr:2} is computed from profiles of broken bonds in the network obtained by accumulating scission events only over a stretch window $(\lambda, \lambda+\Delta \lambda)$ (with $\Delta \lambda = 0.4$) (instead of accumulating since the beginning of the simulation as shown in Fig.~\ref{fgr:A4}). We also use a slightly larger bin size  $\Delta x = 20$ in order to collect enough events, although the curve shown in Fig.~3 (averaged over three samples) remains noisy due to the relatively small number of scission events in our samples.

%%%%%%%%%%%%%%%%%%%%%%%%
\section{Response to individual scission events}\label{app:single-bb}

%%%%%%%%%%%%%%%%%%%%%%%%%%%%%%% FIGURE A5 %%%%%%%%%%%%%%%%%%%%%%%%%%%%%%%%%%%%%%%%%%%%%% 
\begin{figure*}[t]
  \includegraphics[width=1.0\textwidth]{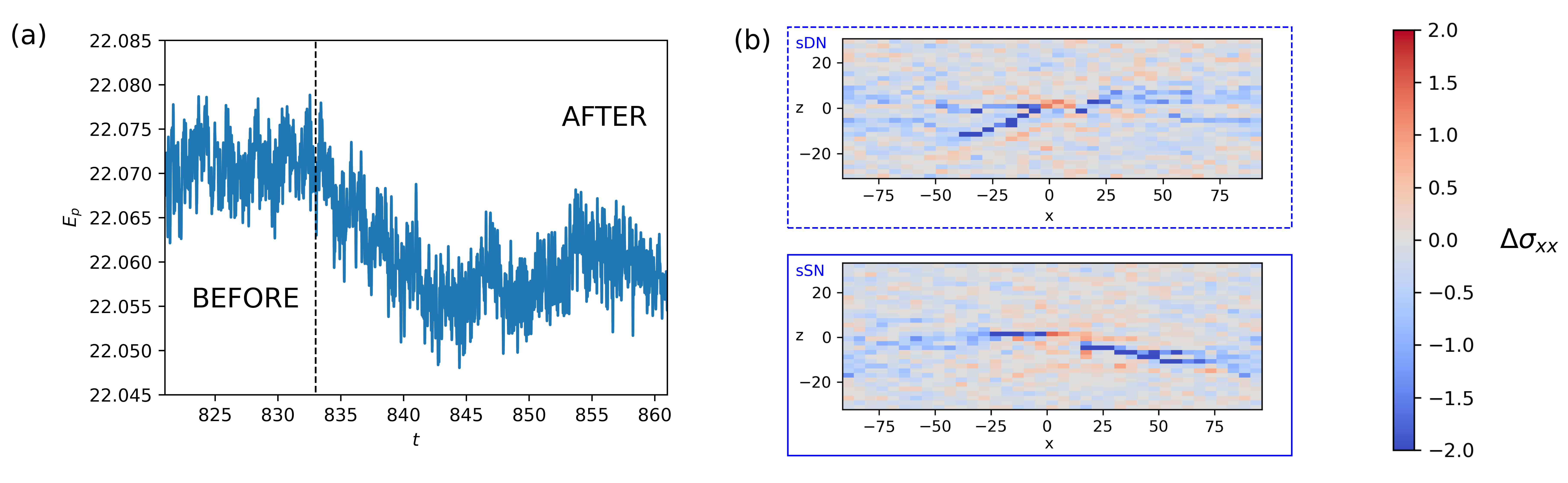}
    \caption{Protocol to measure the average response to single bond breaking events. 
    (a) Total potential energy of a single network sample as a function of time (in LJ units).
    (b) Examples of stress responses induced by a single bond breaking event (not averaging over several bond scission events) in the sacrificial network inside a DN (sDN) (top pane) and in a sacrificial single network (sSN) (bottom panel). The stress map is obtained by integrating along $y$ over the sample thickness and averaging in the (x, z) plane using bins of size $dx = 5\sigma$, $dz = 2\sigma$.}
  \label{fgr:A5}
\end{figure*}
%%%%%%%%%%%%%%%%%%%%%%%%%%%%%%%%%%%%%%%%%%%%%%%%%%%%%%%%%%%%%%%%%%%%%%%%%%%%%%%%%%%%%%%%%%%%%%%%

Fig.~\ref{fgr:3} of the main text depicts the average stress responses to a single bond breaking event in regime (i), (ii)a and (ii)b. 
At the macroscopic scale, a single bond breaking event leads to energy dissipation (as shown by the drop in potential energy in Fig.~\ref{fgr:A5}(a)).

Average stress change values $\langle \Delta \sigma_{xx}\rangle$ following a single bond scission event (spatial average of the stress maps shown in Fig.~\ref{fgr:3}) are measured for the different networks in the different regimes, for a bond breaking event occurring in the sacrificial network (Table~\ref{table:A1}) and in the matrix network (Table~\ref{table:A2}). 

\begin{table}[h]
\begin{tabular}{|c|c|c|c|c|}
\hline
Reg. & sDN & mDN & sSN & SN \\
\hline
(i) & $ - 0.12 \pm 0.16 $  & $ - 0.01 \pm 0.04$    & $ - 0.10 \pm 0.14 $  &  $ - 0.12 \pm 0.96 $ \\
\hline
(ii)a & $ - 0.11 \pm 0.29 $  & $ -0.08 \pm 0.28 $ &  $ - 0.11 \pm 0.22  $ & - \\
\hline
(ii)b & $ - 0.11 \pm 0.21$  & $ -0.12 \pm 0.12 $   &  -  & - \\
\hline
(iii) & $ -0.13 \pm 0.38 $   & $ -0.32 \pm 0.18 $  & -  & - \\
\hline
\end{tabular}
\centering

\caption{Average stress change following a single scission event in the sacrificial network}
\label{table:A1}
\end{table}

\begin{table}[h]

\begin{tabular}{|c|c|c|c|c|}
\hline
Reg. & sDN & mDN  \\
\hline
(ii)b & $ - 0.05 \pm 0.14$  & $ -0.17 \pm 0.18 $   \\
\hline
(iii) & $ -0.17 \pm 0.97 $   & $ -0.48 \pm 0.17 $  \\
\hline
\end{tabular}
\centering

\caption{Average stress change following a single scission event in the matrix network}
\label{table:A2}
\end{table}

%\subsection{Simulation protocol}
Since thermal motion induces significant particle displacement and stress and energy fluctuations even in the absence of bond breaking, we average the atoms’ positions and the stress over several instantaneous snapshots sampled before and after the bond breaking event (see
Fig.~\ref{fgr:A5}(a)) in order to obtain the response (change in stress) by comparing typical configurations. 
Network configurations are saved every $\tau - 2\tau$ (depending upon the rate of bond breaking),
before and after the bond breaking event. Typical configurations before and after a bond breaking event are obtained by averaging over at least 150 snapshots. 

The $xx$ component of the stress change is computed as : $\Delta \sigma_{xx} = \sigma^\mathrm{after}_{xx}  - \sigma^\mathrm{before}_{xx}$.
The (thermally averaged) stress change in response to a single BB event is depicted for a sDN (see Fig.~\ref{fgr:A5}(b) top panel) and a sSN (see Fig.~\ref{fgr:A5}(b), bottom panel). 

Maps shown in Fig.~\ref{fgr:A5} are obtained by integrating along $y$ over the sample thickness and averaging in the (x, z) plane using bins of size $dx = 5\sigma$, $dz = 2\sigma$.

%\subsection{Data analysis}
In order to obtain the average maps shown in Fig.~\ref{fgr:3} of the main text, we then average the response over several bond breaking events.
% Number of events for each BB type in each regime for each network type
Average responses are computed using 
24 scission events in regime (i),
32 events in regime((ii)a,
21 events in regime (ii)b
and 31 event for the sSN in in regime (i).

% CITE SI8
The response to bond scission in the matrix network is shown in the Supplementary Information (Fig.~SI8), obtained by averaging over 15 scission events in regime (ii)b and 32 events in regime (iii).

% CITE SI9
The response to bond scission in single network with $\lambda_0=1.00$ (unswollen) is shown in the Supplementary Information (Fig.~SI9), obtained by averaging over 25 scission events in regime (i). Interestingly, the relaxation of strands in the SN is more aligned to the stretching direction compared to the isotropically pre-stretched samples (sSN and sDN). 

% CITE SI10
We checked that the response to bond scission measured with our method does not depend on system size, by comparing the response of two different system sizes (Fig.~SI10).

% Stress as a function of r
The evolution of stress as a function of the distance to bond breaking $r$ are obtained by azimuthally averaging data in a slice of thickness $dx = \sigma$ centered around $x = 0 $ and parallel to the $(y,O, z)$ plane.

\clearpage

\onecolumngrid

%%%%%%%%%%%%%%%%%%%%%%%%%%%%%%%%%%%%%%%%%%%%%%%%%%%%%%%%%%%%%%%%%%%%%%%%%%%%%%%%%%%%%%%%%%%%%%%%
%%%%%%%%%%%%%%%%%%%%%%%%%%%%%%%%%%%%%%%%%%%%%%%%%%%%%%%%%%%%%%%%%%%%%%%%%%%%%%%%%%%%%%%%%%%%%%%%
%%%%%%%%%%%%%%%%%%%%%%%%%%%%%%% APPENDIX TO FIG 5 %%%%%%%%%%%%%%%%%%%%%%%%%%%%%%%%%%%%%%%%%%%%%% 
\section{Non-affine displacement and broken bonds around a damage islands} \label{app:notch}

%%%%%%%%%%%%%%%%%%%%%%%%%%%%%%% FIGURE A6 %%%%%%%%%%%%%%%%%%%%%%%%%%%%%%%%%%%%%%%%%%%%%%
\begin{figure*}[h]
  \includegraphics[width=\textwidth]{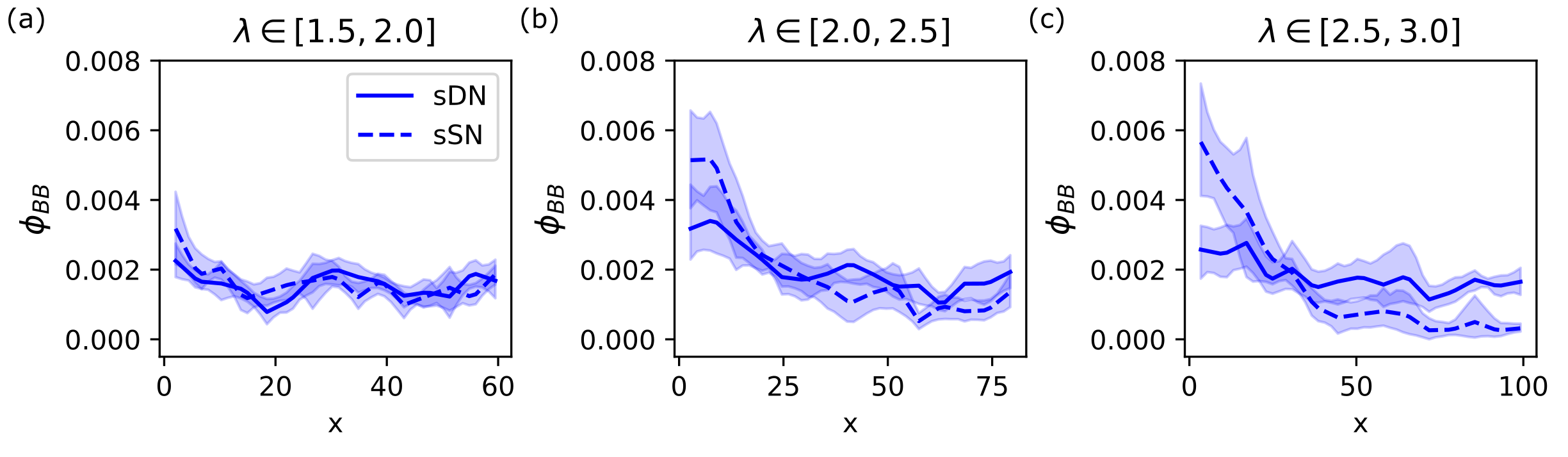}
  \caption{ Fraction of broken bonds ($\phi_{BB}$) as a function of the distance from the pre-damage center (located at $x = 0$) for the pre-damaged SSN and DN systems. The quantity $\phi_{BB}$ is evaluated within the stretch window indicated in each panel. }
  \label{fgr:A6}
\end{figure*}
%%%%%%%%%%%%%%%%%%%%%%%%%%%%%%%%%%%%%%%%%%%%%%%%%%%%%%%%%%%%%%%%%%%%%%%%%%%%%%%%%%%%%%%%%%%%%%%%

The non-affine displacement shown in panels (f) and (i) of Fig.~\ref{fgr:4} was computed for the sacrificial network, excluding dangling-end strands, whose contribution could otherwise bias the interpretation of the results. The non-affine displacement was first evaluated on a per-atom basis, accounting for the non-affine motion accumulated over the analyzed stretch window. The resulting values were then binned into 20 radial intervals according to the distance from the center of the pre-damage region ($r = 0$). Radial profiles were obtained by averaging over seven independent samples. The shaded error bands represent the standard deviation arising from the scatter among the individual sample profiles.
The profile of bond-breaking events as a function of the absolute position parallel to the stretch direction ($x$), shown in Figure~\ref{fgr:A6}, was computed from the spatial locations of bond breakages, as illustrated in panels (d), (e), (g), and (h) of Fig.~\ref{fgr:4}. These profiles were subsequently averaged over seven samples.

\bibliography{multinet}

\clearpage

\section*{Supplementary Information for : Revealing the origin of spatio-temporal damage evolution in double polymer networks}

\setcounter{figure}{0} % Réinitialiser le compteur
\renewcommand{\thefigure}{SI\arabic{figure}} % Changer le format de numérotation


\section{Strand contour length}

\begin{figure}[h]
  \includegraphics[width=0.5\columnwidth]{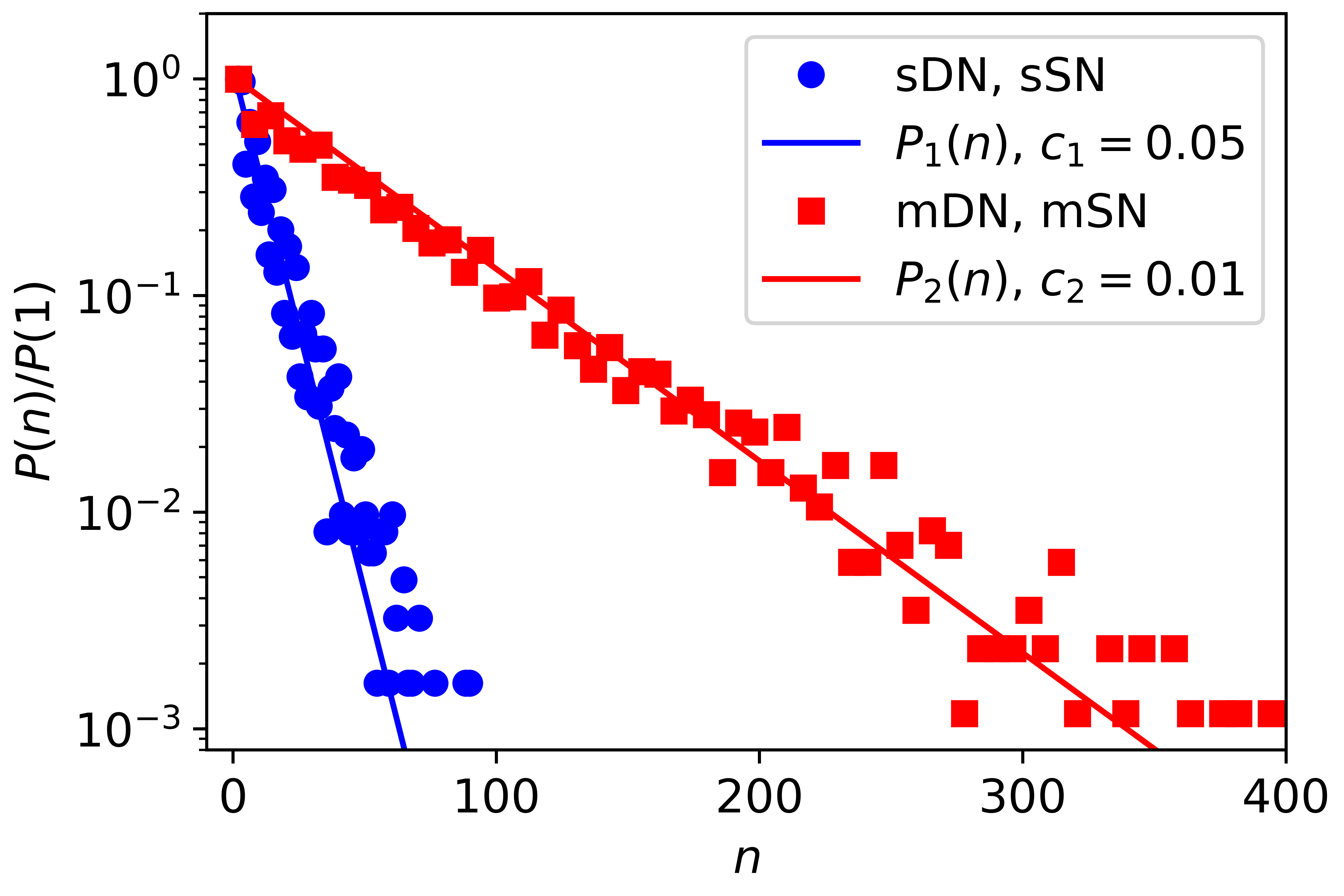}
  \caption{Distributions of strands contour lengths in the sacrificial network and in the matrix network inside the DN (sDN and mDN, respectively), which remain unchanged when deleting the second network to form a single network (sSN and mSN, respectively). Lines correspond to $P(n) = (1- 1/\langle N_i \rangle)^(n-1) $ with $\langle N_1 \rangle = 0.5 (1-c_1)/c_1$ and $\langle N_2 \rangle$ the average strand lengths in the sacrificial and matrix networks respectively (no fitting  parameter).}
  \label{fgr:SI1}
\end{figure}

Fig.~\ref{fgr:SI1} depicts the strand contour length distribution.
The average strand length in the first network (blue) is smaller than in the second (matrix) network (red), which is synthesized with a lower crosslink density ($c_1 = 5 c_2$).

%%%%%%%%%%%%%%%%%%%%%%%%%%%%%%%%%%%FIGURE SI1 %%%%%%%%%%%%%%%%%%%%%%%%%%
%%%%%%%%%%%%%%%%%%%%%%%%%%%%%%%%%%%%%%%%%%%%%%%%%%%%%%%%%%%%%%%%%%%%%%%%%%%%%%%%%%%%%%%%%%%%%%%%%%%%%%%%%%%%%%%%%%%%

\section{Uniaxial deformation simulations}

\subsection{Different contributions to the macroscopic stress}

%%%%%%%%%%%%%%%%%%%%%%%%%%%%%%%%%%%%%%%%%%%%%%%%%%%%%%%%%%%%%%%%%%%%%%%%%%%%%%%%%%%%%%%%%%%%%%%%
%%%%%%%%%%%%% ENHANCEMENT FACTOR, REGIME 2 2 inter-network coupling %%%%%%%%%%%%%%%%%%%%%%%%%%%%
%%%%%%%%%%%%%%%%%%%%%%%%%%%%%%%%%%%%%%%%%%%%%%%%%%%%%%%%%%%%%%%%%%%%%%%%%%%%%%%%%%%%%%%%%%%%%%%%

\begin{figure}[h]
  \includegraphics[width=0.6\columnwidth]{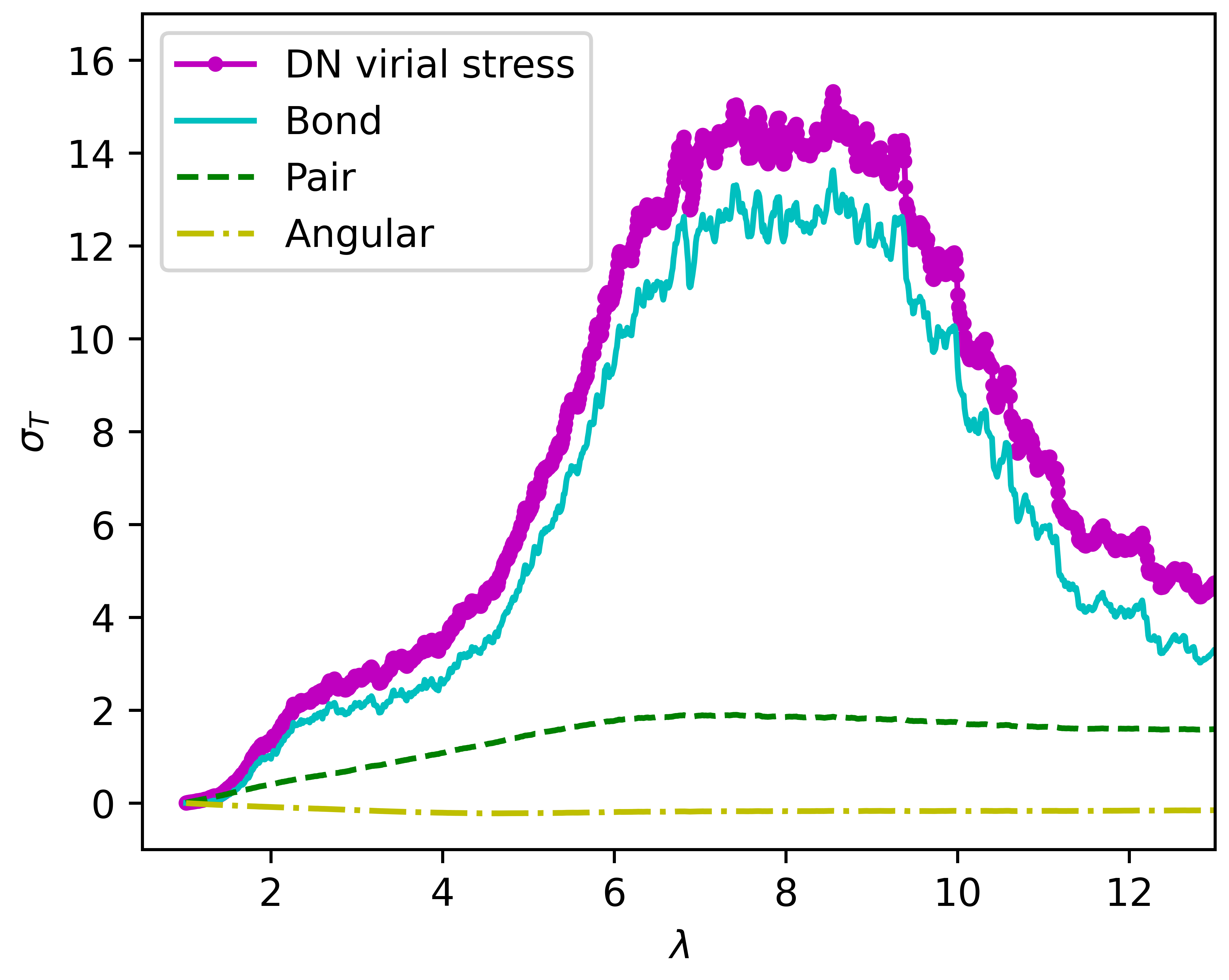}
  \caption{Stress strain curve for the double network with $\lambda_0 = 1.75$ (magenta dots and solid line) and for the different contributions to the stress arising from the various terms in the interaction potential: pair potential (green star symbols), bond potential (cyan big circles) and angular potential (yellow small circles).}
  \label{fgr:SI2}
\end{figure}

Fig.~\ref{fgr:SI2} depicts the different contributions of the interaction potential to the stress-strain curve, and one can see that the macroscopic stress response is dominated by the bond interactions.

\subsection{Stress enhancement factor and inter-network coupling}

\begin{figure}[h]
    \includegraphics[width=0.6\columnwidth]{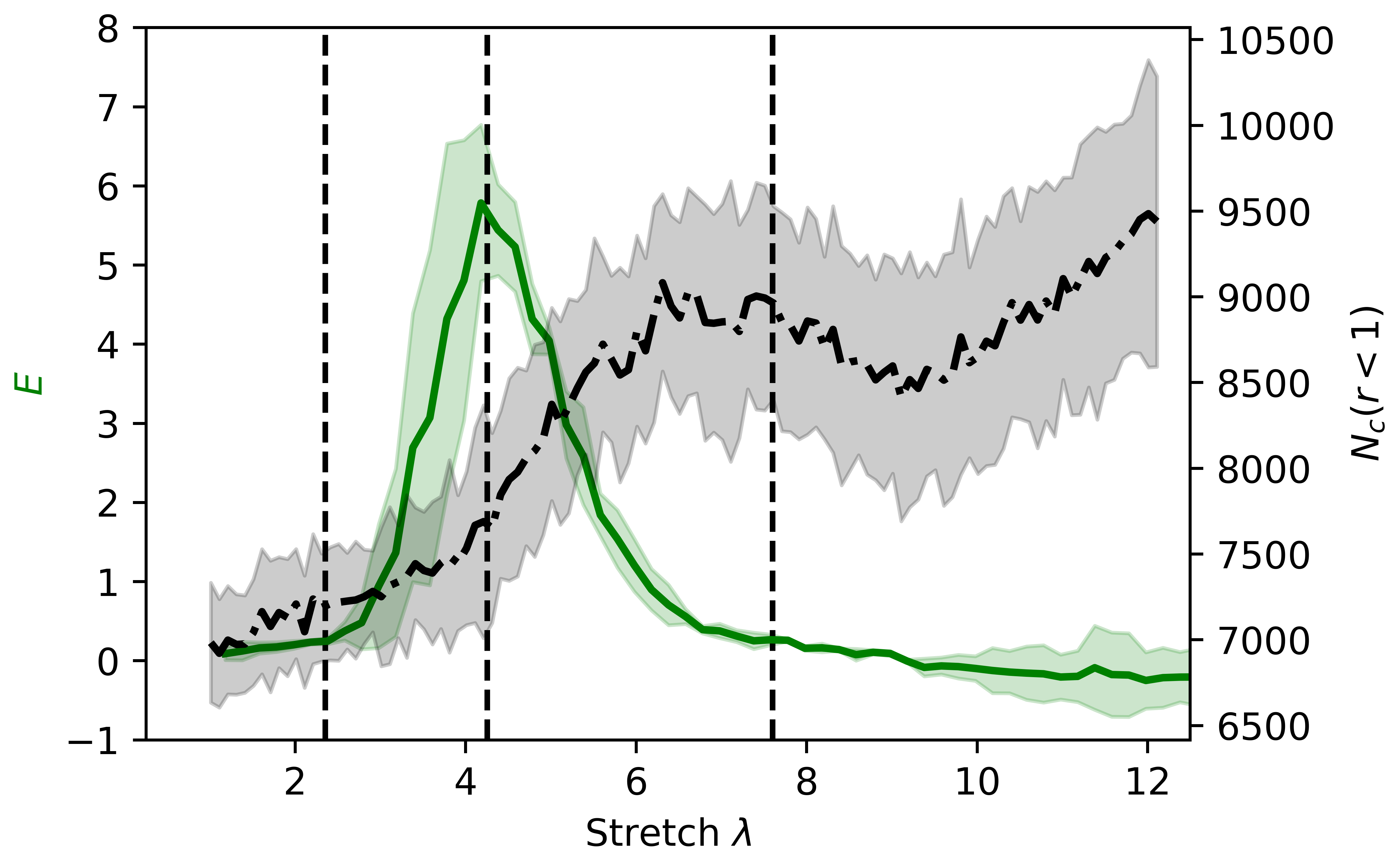}
  \caption{Enhancement factor (green solid line) computed by comparing the response of double networks (DN) and single networks (sSN and mSN) with a pre-stretch $\lambda_0=2.00$. Number of inter-network contacts $N_c$ within a radius $r<r_c$ with $r_c=1$ (black dot-dashed line). Data is averaged over three independent samples.}
  \label{fgr:SI3}
\end{figure}

The stress enhancement factor can be computed by comparing the response to uniaxial deformation of the DN to the response of individual single networks (sSN and mSN, see Fig.~1(b) in the main text):
\begin{equation}
    E = \frac{\sigma_\mathrm{DN}}{\left ( \sigma_\mathrm{sSN} + \sigma_\mathrm{mSN} \right )} - 1
\end{equation}
which starts to grow in regime (ii)a and exhibits a maximum at the beginning of regime (ii)b, when the matrix starts to be loaded (Fig.~\ref{fgr:SI3}(a), green solid line).
This stress enhancement arises due to the steric interactions between the two networks, which also start to grow in regime (ii) and reach a maximum just before the macroscopic fracture start to propagate (regime (iii)), as indicated by the number of inter-network contact within a radius $r<r_c=1$ (black dot-dashed line in Fig.~\ref{fgr:SI3}).

\subsection{Bond breaking dynamics}

\begin{figure}[t]
  \includegraphics[width=0.5\columnwidth]{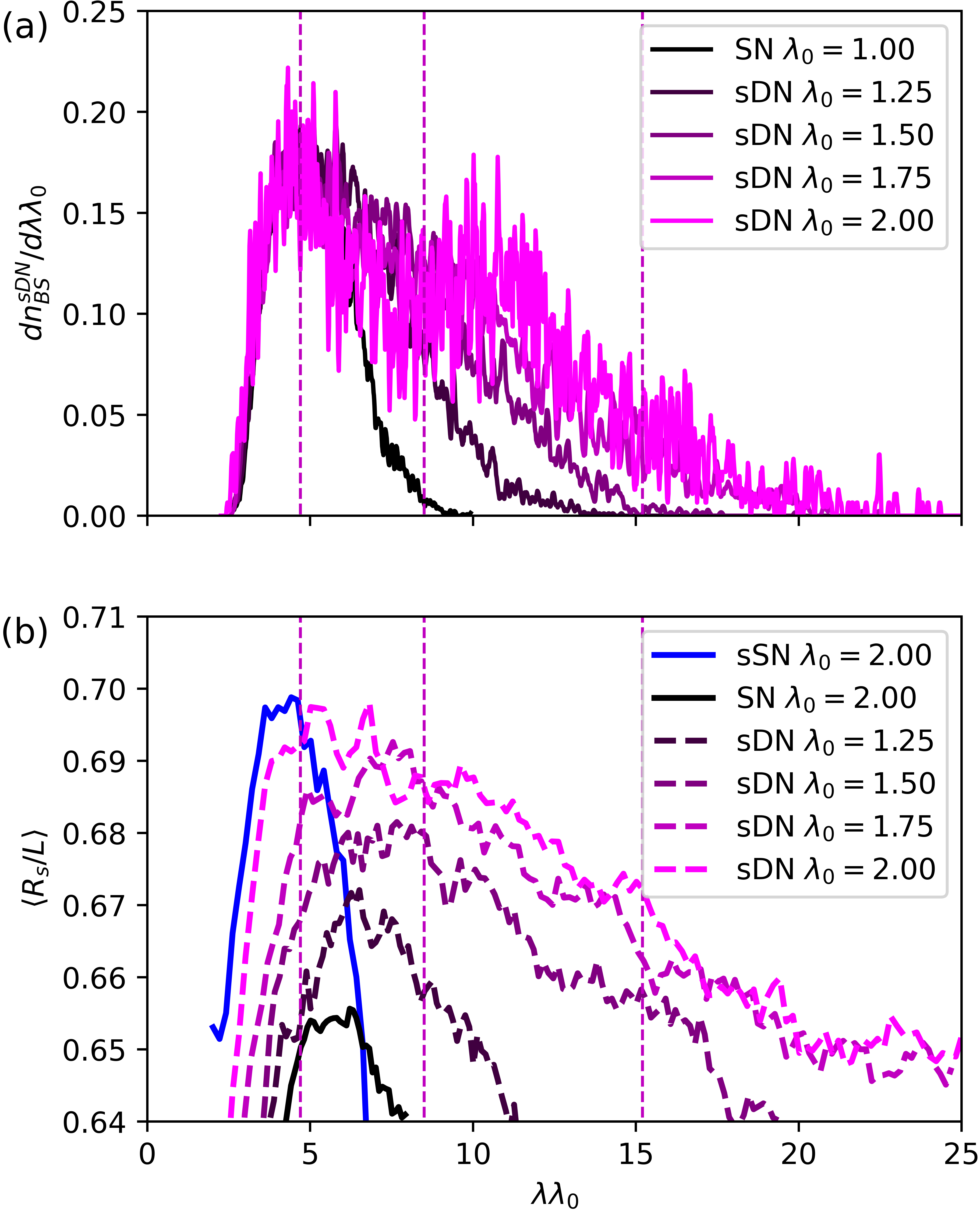}
  \caption{ (a) Rupture rate in the sacrificial network as a function of rescaled stretch $\lambda \lambda_0$ (derivative of the data of Fig.~\ref{fgr:1}(b) with respect to $\lambda \lambda_0$).
  (b) Average strand stretch in the sacrificial network $\langle R_s / L \rangle$ as a function of rescaled stretch $\lambda \lambda_0$. Dashed vertical magenta lines indicate the boundaries of the different regimes for  $\lambda_0 = 2.00$. }
  \label{fgr:SI4}
\end{figure}

Although the two-steps dynamics (as evidenced by Tauber et al.~\cite{tauber2021sharing}) can already be seen by looking closely at Fig.~2(a) in the main text, it appears more clearly when plotting the derivative of the fraction of broken strands with respect to stretch.  
In particular, one can see that the two-steps dynamics is even present for the damage occurring in the sacrificial network (Fig.~\ref{fgr:SI4}(a)), which exhibits a second maximum of rupture rate in regime (ii)b.

\subsection{Strand stretch}
Fig.~\ref{fgr:SI4}(b) depicts the average strand stretch in the sacrificial network for different values of the pre-stretch $\lambda_0$.
While sacrificial strands quickly relax in single networks (sSN and SN), they remain stretched during a larger stretch window as $\lambda_0$ increases, thus leading to more progressive damage.

\subsection{Individual stress-strain curves and toughness measurement}

\begin{figure}[t]
    \includegraphics[width=0.5\columnwidth]{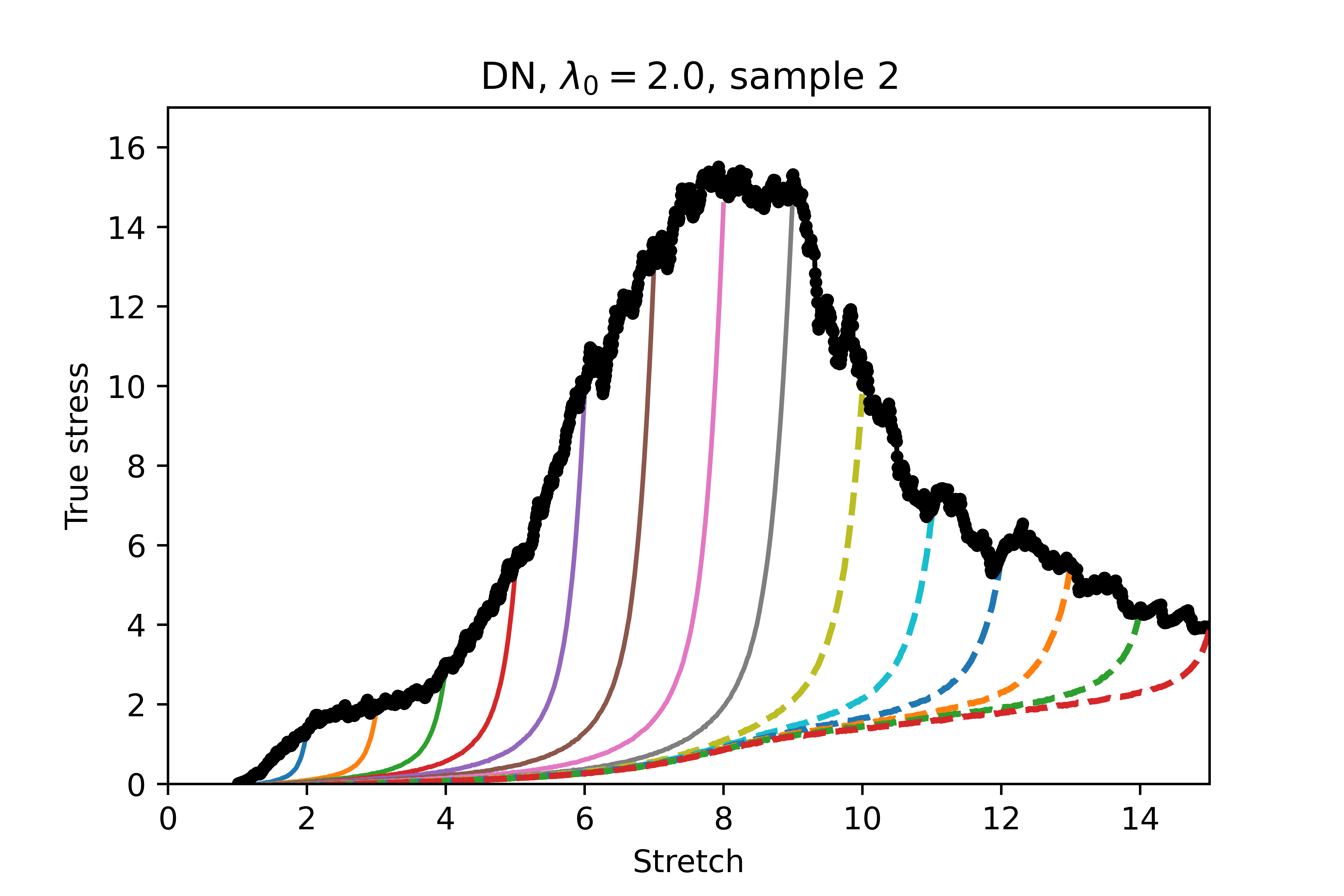}
  \caption{Stress-strain curves for a single double network sample with $\lambda_0 = 2.00$. The stress-strain curve obtained during uniaxial loading of an undamaged sample is shown with the thick black line. Thin colored lines depict the stress-strain curves obtained by unloading the damaged sample before (solid) and after (dashed) the stress peak.}
  \label{fgr:SI5}
\end{figure}

Fig.~\ref{fgr:SI5} depicts an example of stress-stretch curve obtained for an individual double network sample with $\lambda_0=2.00$,
in a uniaxial loading protocol (thick solid black line)
together with the stress-stretch curves obtained in a uniaxial deloading protocol (thin colored lines).

The energy dissipated up to a stretch $\lambda$ is measured by subtracting the area of the deloading stress-stretch curve (recovered elastic energy) to the area of the loading curve :
$U_{diss} = A_\mathrm{load}(\lambda)-A_\mathrm{deload}(\lambda)$.

The toughness is obtained by taking $\lambda = \lambda_\mathrm{max}$, with $\lambda_\mathrm{max}$ the stretch at which the stress is maximum. Since the maximum is not sharp (and rather forms a plateau), the average toughness is obtained by taking several values in the maximum stress plateau (e.g., pink and grey line in Fig.~SI5), and then averaging over independent samples.

%%%%%%%%%%%%%%%%%%%%%%%%%%%%%%%%%%%%%%%%%%%%%%%%%%%%%%%%%%%%%%
\subsection{Bond breaking spatial dynamics}

Fig.~\ref{fgr:SI5} depicts profiles of density fluctuations, $\rho(x)-\rho_0$, along the stretching direction $x$ together with cumulated broken bonds profiles (up to a stretch $\lambda = 4.0$, end of regime (ii)a).
%At the end of regime (ii)a, bond breaking has occurred only in the sacrificial network and has led to density fluctuations (islands) in the sacrificial network that are already visible at the full DN level.

\begin{figure}[t]
  \includegraphics[width=0.5\columnwidth]{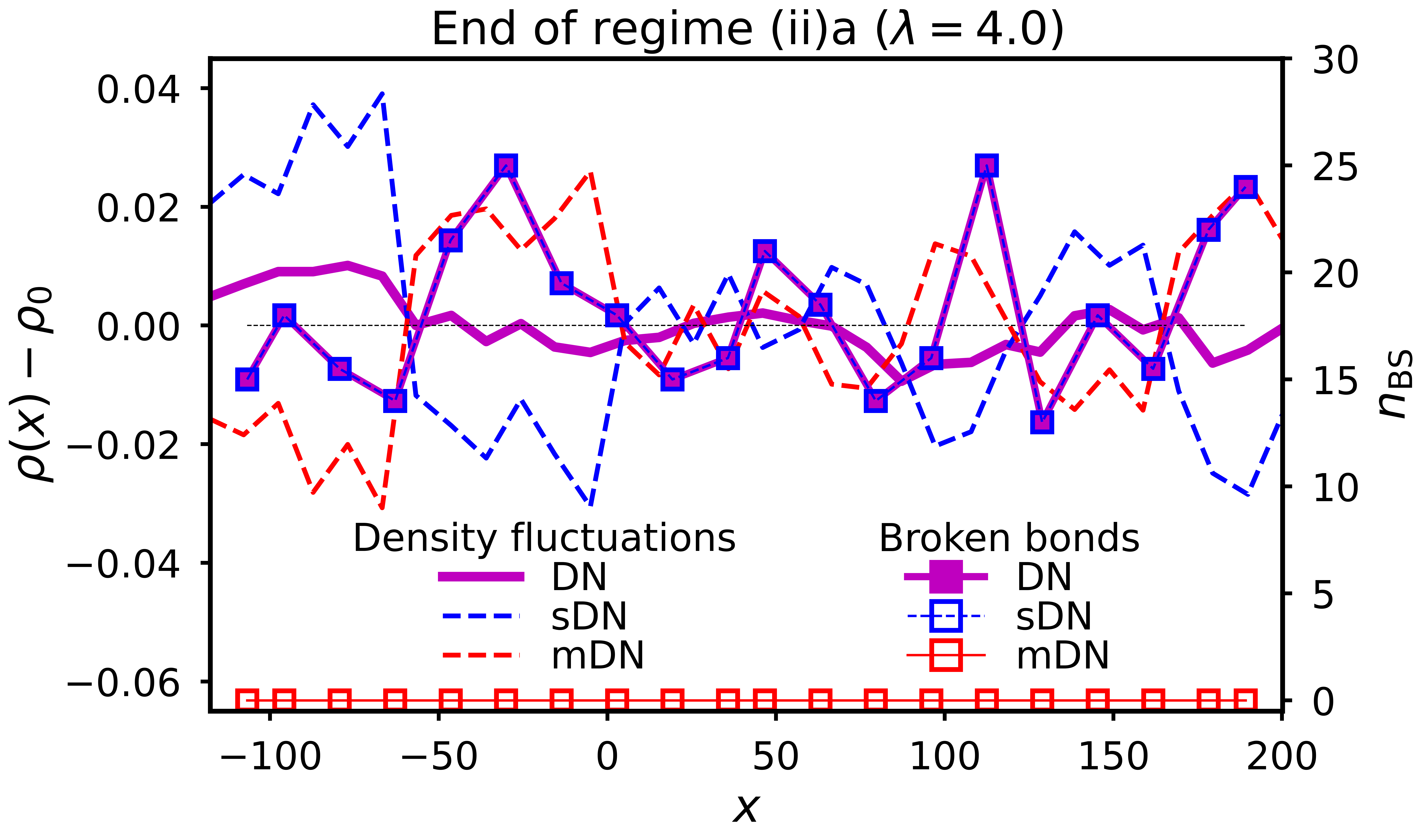}
    \caption{Left axis : Profiles of density fluctuations ($\rho(x)-\rho_0$) for a stretch value $\lambda = 4.0$ (end of regime (ii)a) for a DN with $\lambda_0=2.0$ (solid magenta line), the sacrificial network inside the DN (sDN) (dashed blue line) and the matrix network inside the DN (mDN) (dashed red line).
    Right axis : Corresponding profiles of the number of broken strands along the stretching direction (accumulated damage throughout the simulation) where damage has occured only in the sacrificial network (blue squared symbols.)}
  \label{fgr:SI6}
\end{figure}
%%%%%%%%%%%%%%%%%%%%%%%%%%%%%%%%%%%%%%%%%%%%%%%%%%%%%%%%%%%%%%%%%%%%%%%%%%%%%%%%%%%%%%%%%%%%%%%%%%%

\begin{figure}[t]
  \includegraphics[width=0.5\columnwidth]{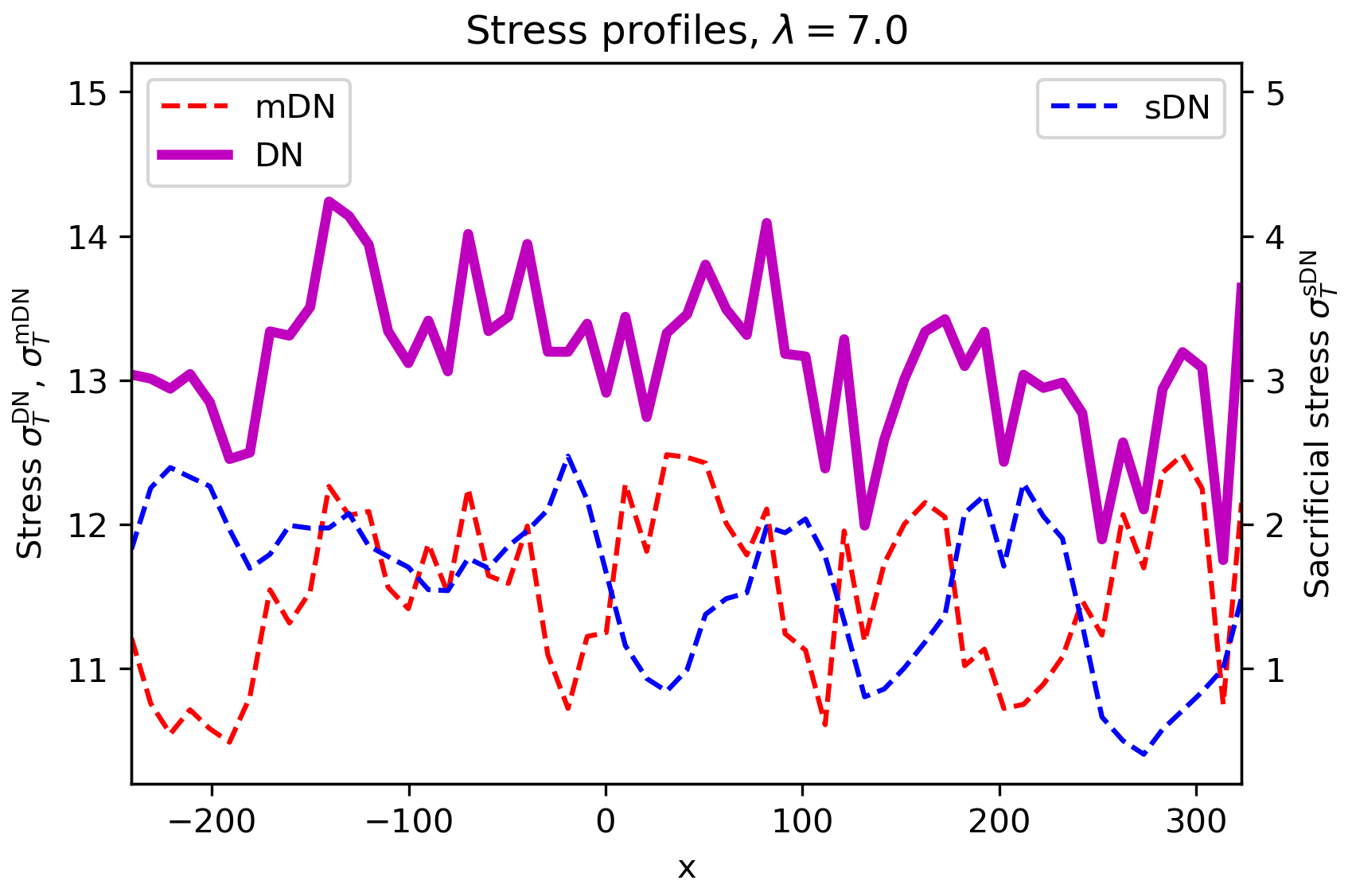}
  \caption{Profiles of true stress along the stretching direction $x$ measured at a stretch value $\lambda = 7.0$. Shown is the stress in the full DN (magenta solid line), in the sacrificial network (blue dashed line) and in the matrix network (red dashed line).}
  \label{fgr:SI7}
\end{figure}

The sacrificial network gets damaged into islands and yields density heterogeneities, which induce a stress concentration onto the matrix network on these specific regions, as shown in Fig.~\ref{fgr:SI7}.

%%%%%%%%%%%%%%%%%%%%%%%%%%%%%%%%%%%%%%%%%%%%%%%%%%%%%%%%%%%%%%%%%%%%%%%%%%%%%%%%%%%%%%%%%%%%%%%%%%%
\subsection{Response of the sacrificial and matrix networks to matrix bond breaking in the different regimes}

\begin{figure}[t]
  \includegraphics[width=\textwidth]{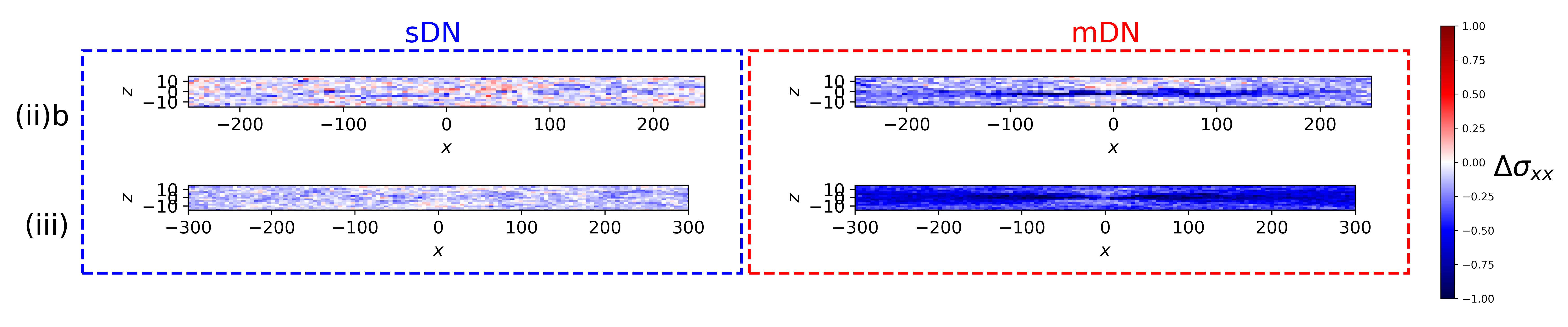}
  \caption{Bond breaking in the matrix network in regimes (ii)b (top) and (iii) (bottom panel). Stress response in the sacrificial network sDN (left) and in the matrix network, mDN (right).}
  \label{fgr:SI8}
\end{figure}

We show in Fig.\ref{fgr:SI8} the average stress response to a bond breaking event occurring in the matrix network in the DN (averaged over 15 events in regime (ii)b and 32 events in regime (iii)).

%%%%%%%%%%%%%%%%%%%%%%%%%%%%%%
\subsection{Response of the sacrificial and matrix networks to matrix bond breaking in the different regimes}

We show in Fig.\ref{fgr:SI8} the average stress response to a bond breaking event occurring in the matrix network in the DN (averaged over 15 events in regime (ii)b and 32 events in regime (iii)).

\begin{figure}[t]
  \includegraphics[width=\textwidth]{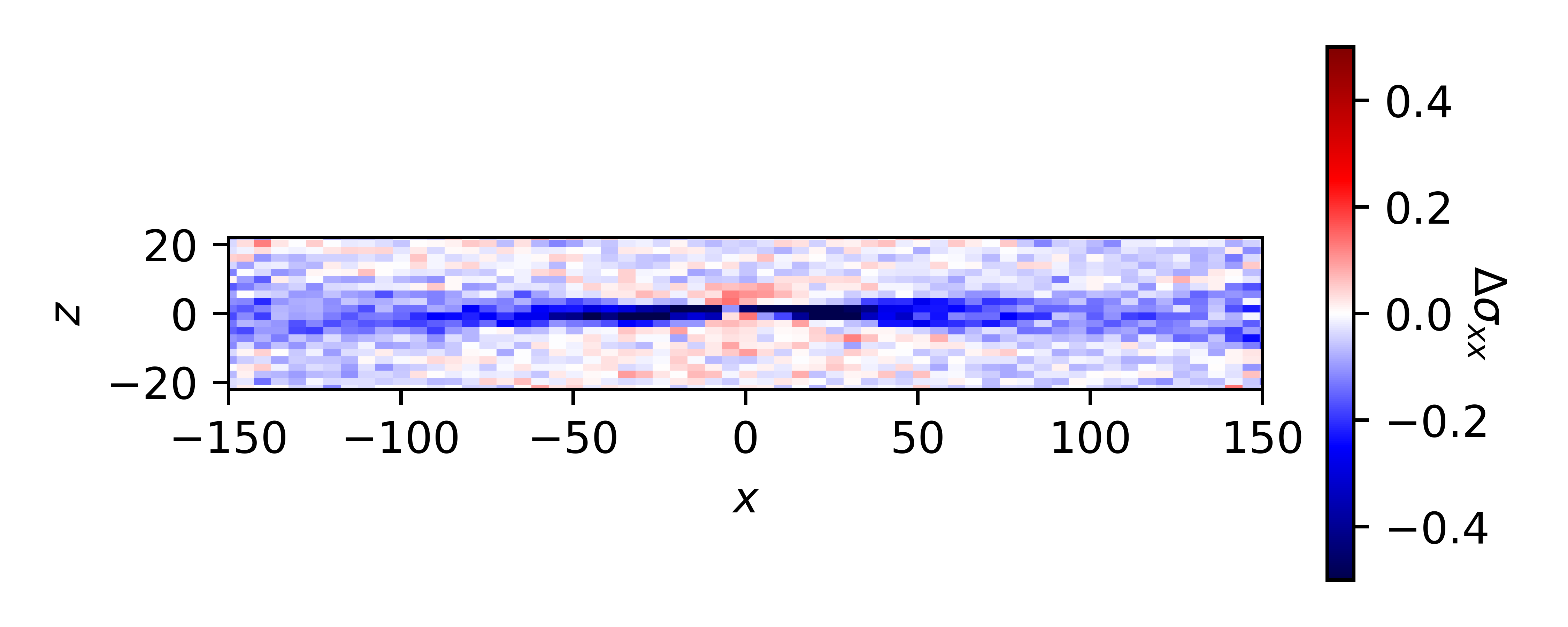}
  \caption{Stress response $\Delta \sigma_{xx}$ to bond breaking in the single network (SN) in regime (i)}
  \label{fgr:SI8b}
\end{figure}

%%%%%%%%%%%%%%%%%%%%%%%%%%%%%%%%%%%%%%%%%%%%%%%%%%%%%%%%%%%%%%%%%%%%%%%%%%%%%%%%%%%%%%%%%%%%%%%%%%%

\subsection{Finite size effects}

In order to check that the range and magnitude of the response to single bond breaking events is not dominated by finite size effects (with our periodic boundary conditions), we repeated the same analysis using a smaller system with $L_0^\mathrm{small}=28 \simeq L_0/3$.

\begin{figure}[t]
  \includegraphics[width=0.7\textwidth]{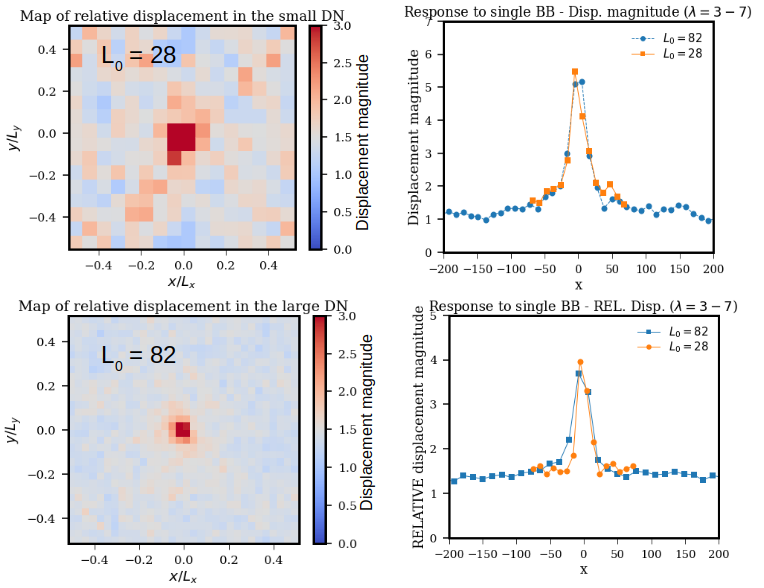}
  \caption{Left : maps of relative displacement magnitude in the small DN ($L_0=28$) (top) and in the large DN ($L_0=82$). Right: Displacement magnitude as a function of the distance $x$ to the bond scission event along the stretching ($x$) direction. Shown are the absolute (top) and relative (bottom) displacement magnitudes.}
  \label{fgr:SI9}
\end{figure}

Fig.~\ref{fgr:SI9} depicts the average absolute (bottom left) and relative (bottom right) displacement magnitude as a function of the distance to bond breaking in response to a single scission event, measured for the same stretch window for the small and big networks (with an average stretch $\lambda = 5.25$, i.e., regime (ii)).
Relative displacements between average configurations before and after the bond breaking event are computed for pairs of particles in a box of size $\ell \times \ell \times \ell$ (with $\ell(\lambda)$ chosen such that there is 10-20 particles in the box).

Results are averaged over about 30 bond breaking events in each case.
We observe no significant difference in the range and magnitude of the displacement field for the two systems, suggesting that the emergent lengthscale of the response is not due to finite size effects.

%%%%%%%%%%%%%%%%%%%%%%%%%%%%%%%%%%%%%%%%%%%%%%%%%%%%%%%%%%%%%%%%%%%%%%%%%%%%%%%%%%%%%%%%%%%%%%%%
%%%%%%%%%%%%%%%%%%%%%%%%%%%%%%%%%%%%%%%%%%%%%%%%%%%%%%%%%%%%%%%%%%%%%%%%%%%%%%%%%%%%%%%%%%%%%%%%
%%%%%%%%%%%%%%%%%%%%%%%%%%%%%%% APPENDIX TO FIG 5 %%%%%%%%%%%%%%%%%%%%%%%%%%%%%%%%%%%%%%%%%%%%%% 
\section{Matrix stress response at small stretch}

\begin{figure}[t]
  \includegraphics[width=0.5\textwidth]{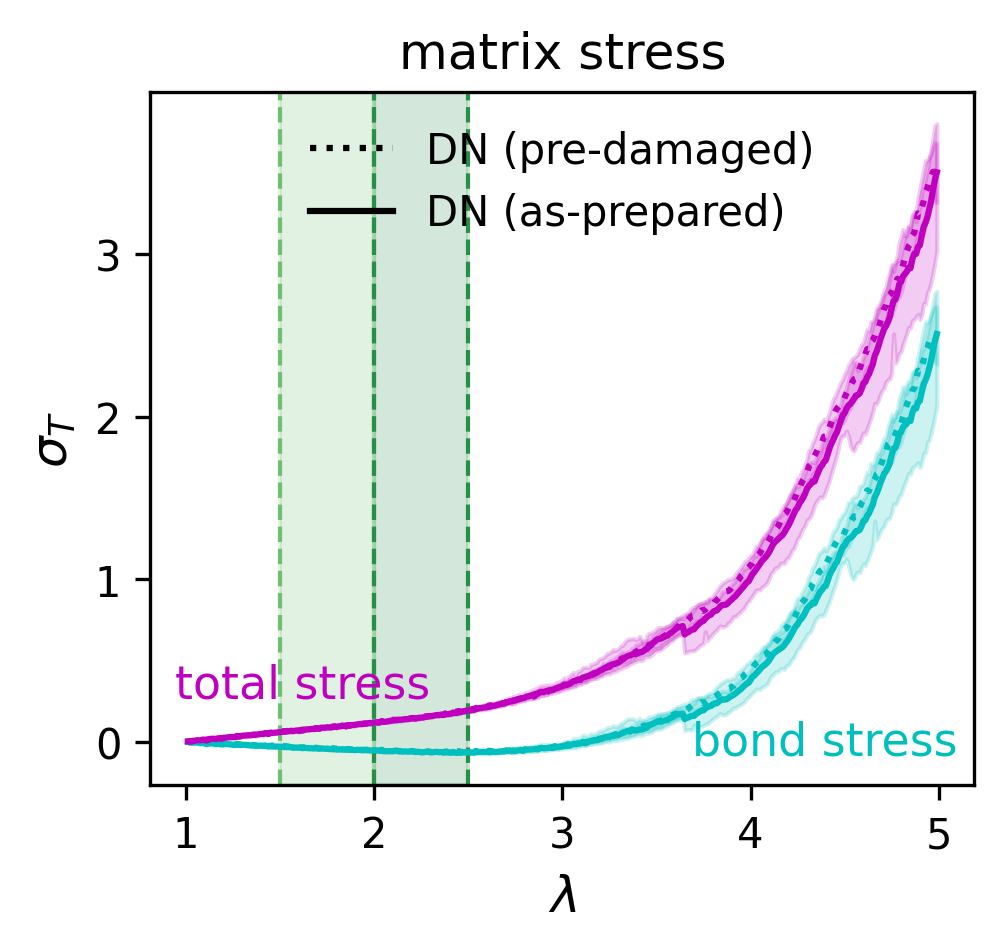}
  \caption{Stress comparison between pre-damaged (dotted line) and as-prepared (not damaged) DN (solid line).
  True stress of the matrix networks are shown, with the total stress indicated in magenta and the bond-related contribution in cyan. }
  \label{fgr:SI10}
\end{figure}

Fig.~\ref{fgr:SI10} depicts the true stress in the matrix network inside a DN (mDN) when a region has been pre-damaged and when the network is intact (as-prepared).

The bond contribution remains negligible in the initial deformation regime for both the pre-damaged and as-prepared DN. Over the entire deformation range, both the total and bond stress components are nearly identical for the pre-damaged and as-prepared systems.

\end{document}